\documentclass[aps,prd,twocolumn,floatfix,preprintnumbers,superscriptaddress,nofootinbib]{revtex4-2}
\usepackage{style}

\begin{document}

\title{Supersymmetric hybrid inflation and current-carrying metastable cosmic strings in $SU(4)_c \times SU(2)_L \times U(1)_R$ }
\author{Adeela Afzal \orcidlink{0000-0003-4439-5342}}
\email{adeelaafzal555@gmail.com}
\affiliation{Department of Physics, Quaid-i-Azam University, Islamabad, 45320, Pakistan}
\author{Maria Mehmood 
\orcidlink{0000-0002-3792-8561}}
\email{mehmood.maria786@gmail.com}
\affiliation{Department of Physics, Quaid-i-Azam University, Islamabad, 45320, Pakistan}
\author{Mansoor Ur Rehman
\orcidlink{0000-0002-1780-1571}}
\email{m.rehman@iu.edu.sa}
\affiliation{Department of Physics, Faculty of Science, Islamic University of Madinah, Madinah 42351, Saudi Arabia}
\author{Qaisar Shafi}
\affiliation{Bartol Research Institute, Department of Physics and Astronomy,
University of Delaware, Newark, DE 19716, USA}

\noaffiliation

\begin{abstract}
We construct a realistic supersymmetric model for superheavy metastable cosmic strings (CSs) that can be investigated in the current pulsar timing array (PTA) experiments. It is based on shifted $\mu$-hybrid inflation in which the symmetry breaking $SU(4)_c \times SU(2)_L \times U(1)_R\rightarrow SU(3)_c\times SU(2)_L  \times U(1)_{B-L}\times U(1)_R$ proceeds along an inflationary trajectory such that the topologically unstable primordial monopoles are inflated away.  {{The breaking of $U(1)_{B-L} \times U(1)_R \rightarrow U(1)_Y$ after inflation ends yields the superheavy metastable CSs carrying the right handed neutrino zero modes that generate the stochastic gravitational wave background (SGWB), which we show is consistent with the current PTA data set and the LIGO O3 run}}. The scalar spectral index $n_s$ and the tensor-to-scalar ratio $r$ are compatible with the Planck 2018  {{and Atacama Cosmology Telescope measurements.}} Both reheating and leptogenesis are briefly discussed.
\end{abstract}

\maketitle

\newpage

\section{Introduction} 
The pulsar timing array experiments have recently shared evidence for the presence of a stochastic gravitational wave background \cite{NANOGrav:2023gor,NANOGrav:2023hvm,EPTA:2023fyk,Reardon:2023gzh,Xu:2023wog}, which may be compatible with the emission of such background by superheavy metastable \cite{Buchmuller:2020lbh,Buchmuller:2021mbb,Lazarides:2023rqf,Buchmuller:2023aus,Antusch:2023zjk,Ahmed:2023rky} or quasi-stable \cite{Lazarides:2023ksx} cosmic strings (CSs).
For a recent discussion of composite topological structures in $SO(10)$, more precisely Spin $(10)$, see ref.~\cite{Lazarides:2023iim}.

In this paper, we describe how the desired metastable strings appear in a realistic supersymmetric inflation model based on the gauge symmetry $G = SU(4)_c \times SU(2)_L \times U(1)_ R$, a well-known subgroup of $SO(10)$. Supersymmetric hybrid inflation models have been extensively studied in \cite{Dvali:1994ms, Copeland:1994vg, Senoguz:2004vu,Rehman:2009nq,Buchmuller:2000zm,Bastero-Gil:2006zpr,urRehman:2006hu,Rehman:2009yj,Shafi:2010jr,Rehman:2010wm,Civiletti:2011qg, Buchmuller:2014epa,Buchmuller:2021dtt,Ahmed:2022rwy, Ahmed:2022wed} and for tribrid inflation see \cite{Masoud:2021prr}. Here we utilize one particular version known as $\mu$-hybrid inflation \cite{Okada:2015vka,Rehman:2017gkm,Okada:2017rbf,Ahmed:2021dvo,Afzal:2022vjx}. Following \cite{Lazarides:2019xai,Lazarides:2023iim}, the breaking of $SU(4)_ c$ to $SU(3)_c \times U(1)_{B-L}$ at a scale close to $M_\text{GUT}$ produces ‘red’ monopoles, that we show are inflated away by employing shifted $\mu$-hybrid inflation \cite{Lazarides:2020zof}. The subsequent breaking of $U(1)_{B-L} \times U(1)_R$ to $U(1)_Y$, also close to $M_\text{GUT}$, takes place after the inflationary phase is effectively over.  {{This produces the desired metastable strings, that are current-carrying due to the associated right handed neutrino zero modes, and which generate a stochastic gravitational wave background}}. The strings eventually disappear from the quantum tunneling of the red monopole-antimonopole pairs.

The paper is laid out as follows. In \cref{model} we summarize the salient features of the $SU(4)_c \times SU(2)_L \times U(1)_R$ model, including symmetry breaking and evolution of the gauge couplings. The shifted $\mu$-hybrid inflationary scenario is presented in \cref{shinf}, and in \cref{sugra} we discuss the supergravity corrections with non-minimal K\"ahler potential. The global minimum and symmetry breaking after inflation is discussed in \cref{Globmin}. \Cref{randl} summarizes how reheating and leptogenesis proceed in this model, together with the inflationary predictions. The metastable strings and predictions related to the gravitational waves (GW) are presented in \cref{CSs}. Our conclusions are given in \cref{con}.
\section{Supersymmetric $SU(4)_c \times SU(2)_L \times U(1)_R$ Model}\label{model}
\begin{table}[t]
\caption{\label{assign1} Matter and Higgs superfields together with their decomposition under $SU(4)_c\times  SU(2)_L \times  U(1)_R = 421$, the $SU(3)_c\times  SU(2)_L \times U(1)_Y=$ SM and their $R$ charges.}
\begin{ruledtabular}
\begin{tabular}{ccccc}
Superfields&$421$ & SM     &$q(R)$  \\
\hline  
\hline
$F_{\ i}$& $({4,\ 2,\ 0})$ 
& $Q_{i} (3,\ 2,\ 1/6)$&1/2 \\
&  & $L_i (1,\ 2,\ -1/2)$&  \\
\hline
$F^c_{u\ i}$& $(\overline{4},\ 1,\ -1/2)$ 
&$U^c_{i}(\overline{3},\ 1,\ -2/3)$&1/2  \\
&  &$N^c_i  (1,\ 1,\ 0)$&\\
\hline
$F^c_{d\ i}$& $(\overline{4},\ 1,\  1/2)$ 
&$D^c_{i}(\overline{3},\ 1,\ 1/3)$&1/2  \\
&   & $E^c_i \ ( 1,\ 1,\ 1)$&\\
\hline
\hline
 $H$& $ ({4,\ 1,\ 1/2})$ & $ u_H( 3,\ 1,\ 2/3)$&0\\
&  & $N_H( 1,\ 1,\ \  0)$&\\
\hline
$\overline {H}$& $ (\overline{4},\ 1,\ -1/2)$ & $\bar{u}_H( \overline{3},\ 1,\ \ \ -2/3)$&0  \\
& & $\overline{N}_H({ 1,\ 1},\ \  0)$& \\
\hline
$ S$& $ ({1,\ 1,\ 1})$ & $ ({ 1,\ 1},\ \ \ 0)$&1 \\
\hline
$ \Phi $& $ ({15,\ 1,\ 0})$ & $({1,\ 1},\ 0)  $&0 \\
& & $({ 8,\ 1},\ \  0)$& \\
& & $({ 3,\ 1},\ \ \ 1/3)$& \\
& & $({ \overline{3},\ 1},\ \ -1/3)$& \\
\hline
 $h_u$&$({1,\ 2,\ \ \ 1/2})$ & $\  ({1,\ 2},\ \ \ 1/2)$&0 \\
 \hline
 $h_d$&$({1,\ 2,\ \ -1/2})$ & $\  ({1,\ 2},\ \ \ 1/2)$&0 \\
\end{tabular}
\end{ruledtabular}
\end{table}
The superfields containing standard model (SM) matter and Higgs content and their $U(1)_R$ charges are given in \Cref{assign1}.  
The $U(1)_R$ symmetric superpotential for shifted $\mu$-hybrid inflation is written as,
\begin{align}\label{superpot-shift}
 W& = S \left(\kappa M^2 -\kappa\, \text{Tr}(\Phi^2) -\dfrac{\beta}{M_\star}\, \text{Tr}(\Phi^3)\right)\nonumber \\
 & +\lambda_H\, S \, H\, \overline{H} 
 + \lambda_h\, S \, h_u\, h_d \nonumber \\
 &+ \lambda^{(u,\nu)}_{ij} F_i F^c_{u\ j} h_u
 +\lambda^{(d,e)}_{ij} F_i F^c_{d\ j}h_d \nonumber \\
 & +  \frac{\gamma_{ij}}{M_*}\, (F^c_{u \,i}\, H)(F^c_{u \,j}\, H),
\end{align}
where $S$ is a gauge singlet superfield, $M_*$ is some 
suitable cutoff scale, and $\left(h_u,h_d\right)$ are the higgs doublet superfields. The third line in $W$ contains Yukawa terms for quarks and leptons. The last term in \cref{superpot-shift} corresponds to the Majorana mass term, which plays a crucial role in elucidating the origin of tiny neutrino masses through the utilization of the seesaw mechanism. 

The $SU(4)_c$ symmetry breaks into $SU(3)_c \times U(1)_{B-L}$ by acquiring a nonzero vacuum expectation value (vev) $v$ in the color singlet direction of the adjoint representation $\Phi$, 
\begin{eqnarray}
SU(4)_c\xrightarrow{\langle\Phi\rangle_{(1,1,0)}} SU(3)_c \times U(1)_{\text{B-L}},
\end{eqnarray}
with,
\begin{equation}\label{globalvev}
  \langle\Phi\rangle_{(1,1,0)}=  \dfrac{v}{2\sqrt{6}}   \begin{pmatrix}
   1 & 0 & 0 & 0 \\
   0 & 1 & 0 & 0 \\
   0 & 0 & 1 & 0 \\
   0 & 0 & 0 & -3
   \end{pmatrix}.
\end{equation}
This breaking creates monopoles that carry $B-L$ and color magnetic charge, which are subsequently diluted during inflation. A similar mechanism is discussed in the standard version of $\mu$-hybrid inflation; see Ref.~\cite{Ahmed:2024iyd}.
To further break $U(1)_{B-L} \times U(1)_R$ into $U(1)_Y$ at $v_H$ we consider $H_{(1,1,0)}$ and $\overline{H}_{(1,1,0)}$,
\begin{eqnarray}
U(1)_{B-L} \times U(1)_R \xrightarrow{\langle H\rangle_{(1,1,0)}=\langle\overline{H}\rangle_{(1,1,0)}} U(1)_Y 
\end{eqnarray}
The SM hypercharge is given by, 
\begin{eqnarray}
    Y=\sqrt{\frac{3}{5}}Y_R+\sqrt{\frac{2}{5}}(B-L),
\end{eqnarray}
where $Y_R$ is the hypercharge associated with $U(1)_R$ and we have chosen the normalization of $Y_R$ assuming $U(1)_R \subset SU(2)_R$. 

\begin{figure}[t]
    \centering
    \includegraphics[width=1.0\linewidth]{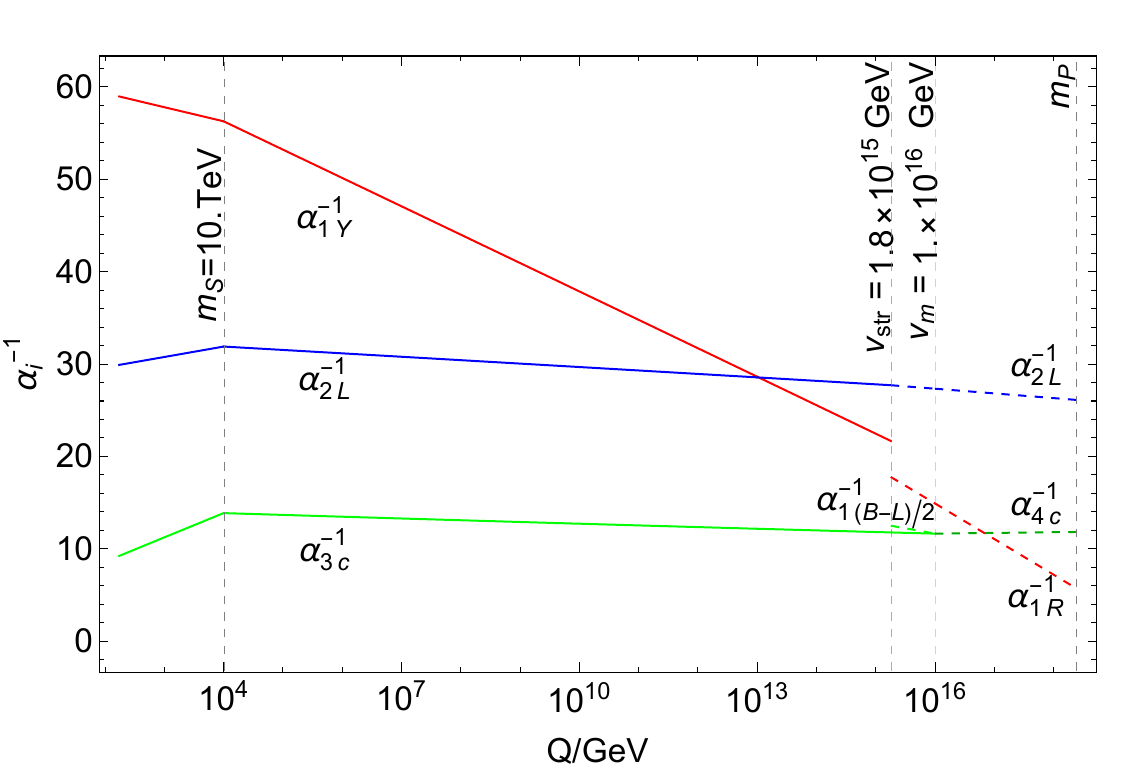}
    \caption{\label{gce}Two loop RG evolution of the gauge couplings for the symmetries $SU(3)_c\times SU(2)_L \times U(1)_Y$, $U(1)_{B-L}$ and $SU(4)_c\times SU(2)_L \times U(1)_R$. Matching conditions at $v_{\text{str}}$ are $\alpha^{-1}_{1Y}=\sqrt{3/5}\ \alpha^{-1}_{1R}+\sqrt{2/5}\ \alpha^{-1}_{1(B-L)}$ and $\alpha^{-1}_{2L}=\alpha^{-1}_{2L}$, and at $v_m$, $\alpha^{-1}_{3c}=\alpha^{-1}_{4c}=\alpha^{-1}_{1(B-L)}$.}
\end{figure}
The two loop gauge coupling evolution for the gauge symmetries $SU(4)_c \times SU(2)_L \times U(1)_R$, $SU(3)_c\times SU(2)_L \times U(1)_R \times U(1)_{B-L}$ and $SU(3)_c\times SU(2)_L \times U(1)_Y$ are shown in \Cref{gce}. We have assumed the degeneracy of supersymmetric particles at $m_S\simeq 10$ TeV. As discussed below, at the MSSM level, a color octet component of the adjoint superfield $\Phi$ and color triplets in $H,\overline{H}$ pair with mass $m_{3/2}\simeq 10$~TeV are present. 
Note that recent LHC data have established a lower bound on the color-octet mass,  $m_{3/2} \gtrsim 2$ TeV \cite{ATLAS:2022pib}. On the other hand, standard cosmology imposes an upper bound of $m_{3/2} \lesssim 2.5$~TeV as discussed in \cite{Ahmed:2022rwy, 
Antusch:2023mxx}. For convenience, we assume that $m_{3/2}$ is degenerate with supersymmetric particles in our analysis. This assumption may be justified in a non-standard cosmological context, as pointed out in \cite{Antusch:2023mxx}, but even if we consider the range $2\ \text{TeV} < m_{3/2} < 2.5$ TeV, our results remain unaffected.
\section{Shifted $\mu$-Hybrid Inflation in Supersymmetric $SU(4)_c \times SU(2)_L \times U(1)_R$ Model}\label{shinf}
In the discussion below, the scalar components of the superfields are denoted by the same symbols as their respective superfields. We express the scalar field $\Phi$ in the $SU(4)_c$ adjoint basis $\Phi=\phi_a\,T^a$, $\text{Tr}[\Phi^2]=\phi^a\,\phi^a/2$,
$\text{Tr}[\Phi^3]=
d^{abc}\,\phi_a\,\phi_b\,\phi_c/4$. Here
$d^{abc}=2\,\text{Tr}[T^a\,\{T^b,\,T^c\}]$
is totally symmetric and $[T^b,\,T^c]=i f^{bcd}\, T^a$
with, $f^{bcd}$ being totally antisymmetric tensors. The indices  $a,b,c$ run from $1$ to $15$, and the repeated indices are summed over. 

The inflation-relevant part of the scalar potential
obtained from the superpotential in \cref{superpot-shift} is given by
\small{
\begin{align}
\label{scalarpotn}
    V &\supset  \bigg| \kappa M^2 - \dfrac{\kappa}{2} \phi_a \phi_a - \dfrac{\beta}{M_{\star}}\dfrac{d^{abc}}{4} \phi_a \phi_b \phi_c + \lambda_H H \overline{H} 
 + \lambda_h h_u h_d\bigg|^2 \notag  \\
    &+ \left| S \right|^2\, \left|\kappa\,\phi_a + \dfrac{3\, \beta}{4\,M_{\star}}\, d^{abc}\,\phi_b\, \phi_c   \right|^2 + \lambda_{H}^2 \left| S \right|^2 ( \left| H  \right|^2 +\left| \overline{H}  \right|^2)   \notag  \\
    &+  \lambda_{h}^2 \left| S \right|^2 ( \left| h_u  \right|^2 +\left| h_d  \right|^2) + V_D,
\end{align}}
where $V_D$ is the D-term potential given by
\begin{align}
V_D \supset g_4^2\,\left(f^{abc}\,\phi_b\, \phi_c^{\star}\right)\,\left(f^{ade}\,\phi_d\, \phi_e^{\star}\right).
\end{align}
 Vanishing of the D-term is achieved by taking, $\phi_a = \phi_{15}\,\delta_{a,15}$, where $\phi_{15}$ is chosen to be a real scalar field. In the D-flat direction, we used an appropriate $R$ transformation to
rotate the $S$ complex field to the real axis, $S=\sigma/\sqrt{2}$, where $\sigma$ is a canonically normalized real scalar field playing the role of inflaton. The supersymmetric global
minimum of the above potential lies at
\begin{figure}[t]
    \centering
    \includegraphics[width=1\linewidth]{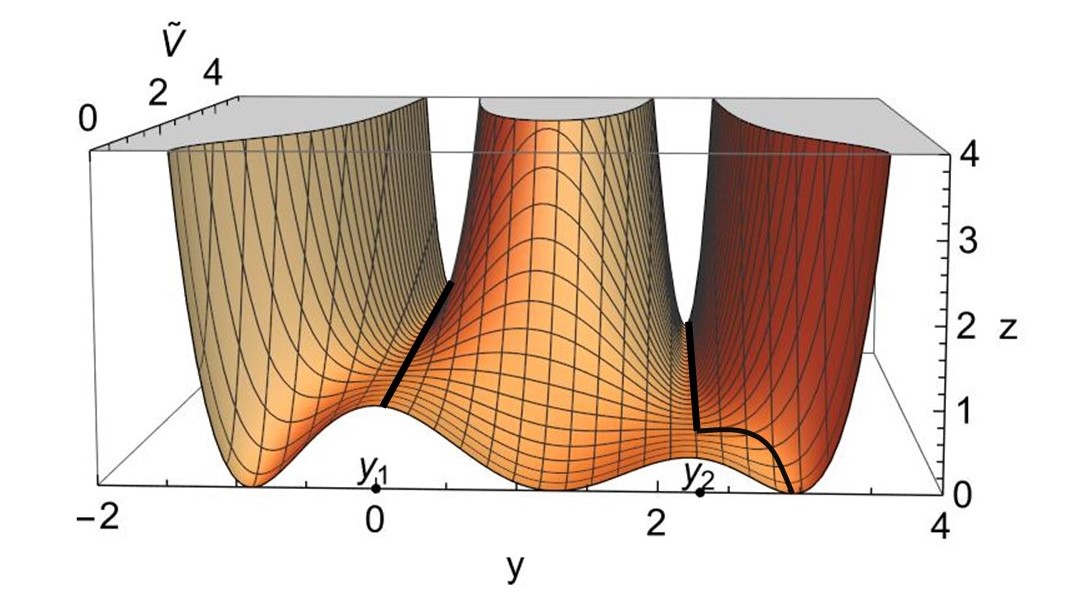}
    \caption{Schematic display of shifted hybrid track for $\xi=0.3$. The two minima at $y_1=0$ and $y_2\simeq 2.3$, correspond to standard and shifted trajectory respectively.}
    \label{fig:scpot}
\end{figure}
\begin{align}\label{globalmin}
     \frac{\kappa}{2}\phi_a^0\,\phi_a^0 +\dfrac{\beta}{4\, M_{\star}}\, d^{abc}\,\phi_a^0\,\phi_b^0\,\phi_c^0  + \lambda_{H} \langle H \overline{H} \rangle= \kappa \,M^2,
\end{align}
where the superscript $``0"$ denotes the field value at its global minimum and,
\begin{eqnarray}
    \langle S \rangle = \langle h_u h_d \rangle = 0, \, \langle H \overline{H} \rangle = v_H^2.
\end{eqnarray}
The acquiring of nonzero vev, $v_H^2$, by $ H \overline{H}$ and the formation of metastable cosmic strings are discussed in later sections. During inflation with $S > M$, the various Higgs fields, namely $h_u\, ,h_d\, ,H\, ,\overline{H}$, acquire heavy masses, leading them to quickly stabilize at zero. Exploiting the $SU(4)_c$ symmetry, the matrix $\Phi^0$ can be transformed into a diagonal form as shown in \cref{globalvev}. Consequently, \cref{globalmin} takes the following form
\begin{equation}\label{Globmininv}
 v_{\Phi}^2  + 4 \gamma_{H} v_H^2 = 4\,M^2  +  \frac{\xi \, v_{\Phi}^3}{2 M},
\end{equation}
where we have used $d_{15,15,15} =-\sqrt{2/3}$,  $\gamma_H = \lambda_H / \kappa$,  $\xi= \beta\,M/\sqrt{3}\,\kappa\,M_{\star}$, and only $\phi^0_{15}= v_{\Phi}/\sqrt{2}\neq 0$.

We can rewrite the scalar potential in \cref{scalarpotn} in terms of the dimensionless fields
\begin{align}
    y=\dfrac{\phi_{15}}{\sqrt{2}\,M}\,\,\,\,\,\,\text{and}\,\,\,\,\,\,z=\dfrac{\sigma}{\sqrt{2}\,M},
\end{align}
with
\begin{align}
    \tilde{V}\equiv\dfrac{V}{\kappa^2\,M^4}=
\left(1-y^2 + \xi\, y^3 \right)^2 +  2\, z^2\, y^2
\left(1- 3\,\xi\, y/2\right)^2.
\end{align}
The schematic display of this potential is shown in \Cref{fig:scpot}. Thus, for a constant value of $z$, the local minima for standard $y_1$ and shifted $y_2$ trajectories are
\begin{align}
    y_1=0\,\,\,\,\,\text{and}\,\,\,\,\, y_2=\dfrac{2}{3\,\xi}.
\end{align}
As explained in \cite{Khalil:2010cp}, the interesting region for the parameter $\xi$ in the current analysis is given by $0.27 \lesssim \xi \lesssim 0.38$.
\begin{table}[t]
\centering {
\begin{tabular}{cc}
\hline\hline {\bf Fields} ~~~~~~& ~~~~~~~~ {\bf Squared Masses} \\
\hline 2 real scalars ~~~~~~~& ~~~~~~~~$\kappa^2 \vert S\vert^2 \pm \kappa^2 M_{\xi}^2 $ \\
\hline 1 Majorana fermion ~~~~~~~& ~~~~~~~~$\kappa^2 \vert S\vert^2 $ \\
\hline 16 real scalars ~~~~~~~& ~~~~~~~~$4 \kappa^2 \vert S\vert^2 \pm 2\kappa^2 M_{\xi}^2 $ \\
\hline 8 Majorana fermions ~~~~~~~& ~~~~~~~~$4 \kappa^2 \vert S\vert^2 $ \\
\hline 8 real scalars ~~~~~~~& ~~~~~~~~$ \lambda_h^2 [S^2\pm M_{\xi}^2 / \gamma_{h} ]$ \\
\hline 4 Majorana fermions ~~~~~~~& ~~~~~~~~$\lambda_h^2S^2$ \\
\hline 16 real scalars ~~~~~~~& ~~~~~~~~$ \lambda_H^2 [ S^2\pm  M_{\xi}^2 / \gamma_H]$ \\
\hline 8 Majorana fermions ~~~~~~~& ~~~~~~~~$ \lambda_H^2S^2$ \\
\hline 6 real scalars ~~~~~~~& ~~~~~~~~$ \frac{2}{3} g^2 v_2^2 $ \\
\hline 6 Dirac fermions ~~~~~~~& ~~~~~~~~$\frac{2}{3} g^2 v_2^2 $ \\
\hline 6 gauge bosons ~~~~~~~& ~~~~~~~~$\frac{2}{3} g^2 v_2^2 $ \\
\hline\hline
\end{tabular}%
} \caption{ Mass spectrum of the shifted hybrid $SU(4)_c$ model as the
system moves along the inflationary trajectory 
$y_2$ ($v_2 = 2\,M\,y_2 = \frac{4\,M}{3\,\xi}$). Here, $M_{\xi}^2 = M^2 (4/27 \xi^2 -1)$, $\gamma_h=\lambda_h/\kappa$ and $\gamma_H = \lambda_H/\kappa$.} \label{Tab:spectrum}
\end{table}
In \Cref{Tab:spectrum}, we summarize the results of the mass spectrum of the model along the shifted inflationary trajectory, $y_2 = 2/(3\xi)$. All these supermultiplets satisfy the supertrace rule $\text{Str}[M^2] = 0$.

The inflationary effective potential with
1-loop radiative correction is given by %
\begin{align}
V_{\rm \tiny{1 loop}} &= \kappa^2 M_{\xi}^4 \big(1+
\frac{\kappa^2}{16\,\pi^2} \left[ F(M_{\xi}^2,x^2)   \right. \big. \nonumber \\
&\big. \left. +8\times 4\,F(2 M_{\xi}^2,2\,x^2) +4\, \gamma^2_h F(m_h^2, x_h^2)  \right. \big. \nonumber \\
&\big. \left.+8\, \gamma^2_H F(M_H^2, x_H^2)\right]\big).
\label{Vloop}
\end{align}
Here
\begin{align} 
F(M_{\xi}^2,x^2) & = 
\frac{1}{4}\left( \left( x^4+1\right) \ln \frac{\left( x^4-1\right) 
}{x^4} +2x^2\ln \frac{x^2+1}{x^2-1} \right. \nonumber \\
&\left.+2\ln \frac{\kappa ^{2}M_{\xi}^{2}x^2}{%
Q^{2}}-3\right),
\end{align}
where $x=|S|/ M_{\xi}$, $M_{\xi}^2 = M^2 (4/27 \xi^2 -1)$, $m_h^2=\gamma_h M_{\xi}^2$, $x_h^2=\gamma_h\, x^2$ with $\gamma_h=\lambda_h/\kappa$ and $M_H^2=\gamma_H M_{\xi}^2$, $x_H^2=\gamma_H\,x^2$ with $\gamma_H = \lambda_H/\kappa$ and
$Q$ is the renormalization scale. To ensure the desired symmetry-breaking vacuum, the coupling constants must satisfy the hierarchy $\lambda_h > \lambda_H \geq \kappa$. The $SU(3)_c$ octet multiplet is massless in the supersymmetric limit as a consequence of both the $R$ and the gauge symmetries. This will not be a problem, as gauge coupling unification is not a prediction of our model. Moreover, these otherwise massless states are anticipated to acquire masses of order $m_{3/2}$ through soft supersymmetry-breaking effects.
\section{SUGRA corrections and nonminimal K\"{a}hler potential}\label{sugra}
We take the following general form of the K\"ahler potential:

\begin{align}
 K=& \vert S \vert^2 + Tr \vert \Phi \vert^2 
+ \vert H \vert^2 + \vert \bar{H}\vert^2 + \vert h_u \vert^2 + \vert h_d \vert^2\nonumber \\
+& \kappa_{S\Phi} \frac{\vert S\vert^2 \, Tr \vert \Phi \vert^2}{m_P^2}
+ \kappa_{S H} \frac{\vert S \vert^2 \vert H \vert^2}{m_P^2}
+ \kappa_{S \bar{H}} \frac{\vert S \vert^2 \vert \bar{H} \vert^2}{m_P^2} \nonumber \\
+& \kappa_{H \Phi} \frac{\vert H \vert^2 \, Tr \vert \Phi \vert^2}{m_P^2} 
+ \kappa_{\bar{H} \Phi} \frac{\vert \bar{H} \vert^2 \, Tr \vert \Phi \vert^2}{m_P^2} 
+ \kappa_{H \bar{H}} \frac{\vert H \vert^2 \vert \bar{H} \vert^2}{m_P^2} \nonumber \\
+& \kappa_S \frac{\vert S\vert^4}{4 m_P^2} 
+ \kappa_{\Phi} \frac{ (Tr \vert \Phi \vert^2)^2}{4 m_P^2} 
+ \kappa_{H} \frac{ \vert H \vert^4}{4 m_P^2} 
+ \kappa_{\bar{H}} \frac{ \vert \bar{H} \vert^4}{4 m_P^2} \nonumber \\
+& \kappa_{SS} \frac{\vert S\vert^6}{6 m_P^4}  + \cdots , %
\label{K} 
\end{align}
where $m_P \simeq 2.4 \times 10^{18}$ GeV is the reduced Planck mass.
Additionally, for the sake of simplicity, the contribution 
of many other terms e.g., of the form 
\begin{eqnarray}
c_2\,[Tr(\Phi^2) + h.c.] + c_3\,[Tr(\Phi^3)/m_P + h.c.] + \cdots
\end{eqnarray}
is ignored. Alternatively, we can 
effectively absorb these extra contributions coming 
from the $\Phi$ superfield into various couplings of the 
above K\"ahler potential as only the $|S|$ field plays an
active role during inflation.
The SUGRA scalar potential is given by
\begin{align}
\label{VF}
V_{F}=e^{K/m_P^{2}}\left(
K_{i\bar{j}}^{-1}D_{z_{i}}WD_{z^{*}_j}W^{*}-3 m_P^{-2}\left| W\right| ^{2}\right),
\end{align}
with $z_{i}$ being the bosonic components of the superfields $z%
_{i}\in \{S,\Phi,H,\bar{H},\cdots\}$ and we have defined
\begin{eqnarray}
D_{z_{i}}W \equiv \frac{\partial W}{\partial z_{i}}+m_P^{-2}\frac{%
\partial K}{\partial z_{i}}W , \,\,\,
K_{i\bar{j}} \equiv \frac{\partial ^{2}K}{\partial z_{i}\partial z_{j}^{*}},
\end{eqnarray}
and $D_{z_{i}^{*}}W^{*}=\left( D_{z_{i}}W\right)^{*}.$ The soft SUSY breaking terms are added in the inflationary potential as:
\begin{equation}\label{soft}
V_\text{soft}=m_{3/2}\big[z_i\frac{\partial W}{\partial z_i}+(A-3)W +h.c.\big],
\end{equation}
where $A$ is the complex coefficient of the trilinear soft-SUSY-breaking terms.
Now along the inflationary trajectory with the D-flat direction 
($\phi_i = \delta_{i,15} \phi_{15}$, $|\bar{H}_a| = |H_a|$)
and using \cref{superpot-shift,Vloop,K,VF,soft}.
we obtain the following form of the full potential:
\begin{eqnarray}
&&V= V_{\rm 1 loop}+\kappa ^{2}M_{\xi}^{4}\left( \left(\frac{4(1-\kappa_{S\Phi})}{9\,(4/27 - \xi^2)} -\kappa_S\,x^2 \right)\right.
\left( \frac{M_{\xi}}{m_P}\right)^{2} \nonumber \\
&& \left. + \left( \frac{4((1-2\kappa_{S\Phi})^2+1+\kappa_{\Phi})}{81\,(4/27 - \xi^2)^2}
 \right. \right. \nonumber \\
&& \left. \left. 
+\frac{4((1-\kappa_{S\Phi})^2-\kappa_{S}(1-2\kappa_{S\Phi}))x^2}{9\,(4/27 - \xi^2)}\right. \right. \nonumber \\
&& \left. \left.+\frac{\gamma _{S}\,x^4}{2}\right) 
\left( \frac{M_{\xi}}{m_P}\right)^{4} +a\,\dfrac{m_{3/2}}{\sqrt{2}\,\kappa\,M_\xi}\,x+\cdots \right) +\cdots\,, 
\label{VT}
\end{eqnarray}
where $x = |S|/ M_{\xi}$ and
$\gamma _{S}=1-\frac{7 \,\kappa _{S}}{2}+2\,\kappa _{S}^{2}-3\,\kappa _{SS}$. Taking into account the range $0 \lesssim |\kappa_{SS}| \lesssim 1$, the parameter $\kappa_{SS}$ does not play a significant role in our numerical computations.
The parameter $a$ depends on $\arg S$ as follows:
\begin{equation}
a = 2\left| 2-A+\frac{A}{2\xi}\right| \cos \left[\arg S+\arg \left(2-A+\frac{A}{2\xi}\right)\right].
\end{equation}

The inflationary slow-roll parameters are given by
\begin{eqnarray}
\epsilon = \frac{m_P^2}{4\,M_{\xi}^2}\left( \frac{\partial_x V}{V} \right)^2, 
\,\,\, \eta = \frac{m_P^2}{2\,M_{\xi}^2} \left( \frac{\partial_x^2 V}{V} \right).
\end{eqnarray}
In the slow-roll approximation, i.e. $(\epsilon$, $|\eta| ) \ll 1$, 
the scalar spectral index $n_s$ and the tensor to scalar ratio $r$ 
are given (to leading order) by
\begin{eqnarray}
n_s \simeq 1 + 2\,\eta -6\,\epsilon, 
\,\,\, r \simeq 16\,\epsilon. 
\end{eqnarray}
The relevant
number of e-folds, $N_0$, before the end of inflation, is
\begin{equation}
N_0 = 2\left( \frac{M_{\xi}}{m_P}\right) ^{2}\int_{x_e}^{x_{0}}\left( \frac{V}{%
\partial _{x}V}\right) dx,
\end{equation}
where $|S_0|=x_0 \, M_{\xi}$ is the field value at 
the pivot scale $k_0=0.05$ Mpc$^{-1}$, and $x_e$ denotes the field value at the 
end of inflation, defined by $|\eta(x_e)| = 1$ (or $x_e = 1$).
During inflation, this scale exits the horizon
at approximately
\begin{equation}
N_{0}=53+\frac{1}{3}\ln \left( \frac{T_{r}}{10^{9}~{\rm GeV}}\right) +\frac{2}{3}%
\ln \left( \frac{V^{1/4}(x_0)}{10^{15}~{\rm GeV}}\right),
\end{equation}
where $T_{r}$ is the reheat temperature and
for the numerical work, we set $T_r = 2\times10^9$ GeV. This
could easily be reduced to lower values if the gravitino problem is considered to be an issue. 
The amplitude of the curvature perturbation is given by
\begin{equation} \label{perturb}
A_s(k_0) = \frac{1}{24\,\pi^2} 
\left. \left( \frac{V/m_P^4}{\epsilon}\right)\right|_{k = k_0},
\end{equation}
where $A_s(k_0) = (2.43\pm 0.11)\times 10^{-9}$ is the Planck 2018 normalization. Note that, for added precision, we include in our calculations the first-order corrections \cite{Senoguz:2008nok} in the slow-roll expansion for the quantities $n_s$, $r$, and $A_s$.


\section{Global Minimum and symmetry breaking after Inflation}\label{Globmin}

 {{We introduce an additional gauge-singlet field $X$, carrying a global $R$-charge of 1, to facilitate the breaking of the symmetry $U(1)_R \times U(1)_{B-L}$ to $U(1)_Y$, which is associated with the formation of a cosmic string (CS) network. The relevant term in the superpotential is given by:
\begin{eqnarray}
W \supset \lambda_X X \left(H \overline{H} - v_H^2 \right),
\end{eqnarray}
where, for simplicity, we leave out the other terms.
In the early universe, prior to observable inflation, the field $X$ is assumed to be stabilized at the origin due to its heavy mass,
$$
m_X^2 \sim \frac{\kappa_X \lambda_X^2 v_H^4}{m_P^2},
$$
which originates from a higher-order term in the K\"ahler potential,
$$
K \supset -\frac{\kappa_X}{4m_P^2} |X|^4,
$$
with $\kappa_X \sim 1$ \cite{Senoguz:2004ky}.
}}

 {{
During inflation, as long as the inflaton field $S$ satisfies $S > M_\xi$, the mass-squared of some components of the waterfall fields is positive:
\begin{equation}
\lambda_H^2 [ S^2  -  M_{\xi}^2/\gamma_H ] > 0,    
\end{equation}
provided $\gamma_H \equiv \lambda_H /\kappa  \geq 1$. This ensures that $|H| = |\overline{H}|$ remains stabilized at zero as the field $\Phi$ rolls toward its minimum after becoming destabilized at $S = M_\xi$.
After the inflationary phase ends, when $S$ drops below $M_\xi/\gamma_H$, the SM-singlet components $|N_H| = |\overline{N}_H|$ of the fields $H, \overline{H}$ become tachyonic and acquire non-zero vacuum expectation values,
\begin{equation}
\langle |N_H| \rangle = \langle |\overline{N}_H| \rangle = v_H.
\end{equation}
This spontaneous symmetry breaking of $U(1)_R \times U(1)_{B-L}$ triggers the formation of the cosmic string network.
The mass of $\theta_H = (\delta N_H + \delta \bar{N}_H)/\sqrt{2}$ component of the waterfall fields, where $\delta N_H = N_H - v_H $, $\delta \bar{N}_H = \bar{N}_H - v_H $, is given by:
\begin{equation}
M_{H} \simeq \sqrt{2} \lambda_X v_H ,
\end{equation}
while the color-triplet components $(u_H,\, \overline{u}_H) \subset (H,\, \overline{H})$ acquire masses of order the gravitino mass, $m_{3/2}$, via soft SUSY-breaking effects \cite{Dvali:1997uq}.

We assume that the formation of the cosmic string network takes place immediately after the end of inflation, near the critical value $S_c \simeq  M_\xi$ with $\gamma_H \simeq 1$. To avoid a secondary phase of inflation triggered by the rolling of the fields $|N_H| = |\overline{N}_H|$, we require that the scalar mass $M_H$ exceeds the Hubble scale during inflation,
$$
M_H > \mathcal{H} \simeq \frac{\kappa\, M_\xi^2}{\sqrt{3}\, m_P} .
$$
This condition ensures that the waterfall fields rapidly settle into their global minima following the end of inflation. Assuming $\lambda_X \sim \kappa$, this leads to a lower bound on the symmetry-breaking scale:
$$
v_H > \frac{M_\xi^2}{\sqrt{6}\, m_P}.
$$
Furthermore, under the assumption $\kappa M_\xi^2 \gg \lambda_X v_H^2$, the vacuum energy during inflation is dominated by the term $\kappa^2 M_\xi^4$, confirming that inflation is primarily driven by the inflaton sector. This hierarchy also implies that $v_H \ll M_\xi$, consistent with the earlier assumption $\lambda_X \sim \kappa$.
}}

\section{ Inflaton Decay and Reheating}\label{randl}
 {{After inflation ends, the system falls towards the SUSY vacuum and performs damped oscillations about it. The oscillating system mainly consists of two complex scalar fields $S$ and $\theta_{\phi} =\delta  \phi$ with the same mass, 
\begin{equation}
m_\text{inf} \simeq \frac{\kappa\, v_{\phi}}{\sqrt{2}}\left(-1+\frac{3\,\xi\, v_{\phi}}{4 M}\right).
\end{equation}
The dominant decay channels of the inflaton (oscillating system) are into a pair of higgsinos ($\widetilde h_u$, $\widetilde h_d$) and higgses ($ h_u$, $ h_d$), each with a decay width, $\Gamma_h$, given by, \cite{Lazarides:1998qx,Okada:2015vka,Afzal:2022vjx}
\begin{eqnarray}\label{gammah}
&&\Gamma_h =\Gamma(S \rightarrow \widetilde h_u\widetilde h_d) = \Gamma(\theta_{\phi} \rightarrow h_u h_d)
=\frac{\lambda_h ^2 }{8 \pi }m_{\text{inf}}. \qquad
\end{eqnarray}
A comparable decay rate exist involving the fields ($ H$, $ \bar{H}$) and their superpartners ($\widetilde H$, $\widetilde{\bar{H}}$), each with a decay width, $\Gamma_H$, given by,
\begin{eqnarray}\label{gammaH}
&&\Gamma_H =\Gamma(S \rightarrow \widetilde H \widetilde{\bar{H}}) = \Gamma(\theta_{\phi} \rightarrow H \bar{H})
=\frac{\lambda_H ^2 }{8 \pi }m_{\text{inf}}, \qquad
\end{eqnarray}
where we take $\lambda_h = 2 \lambda_H$.

Another decay mode for $S$ and $\theta_H$, via the higher-dimensional superpotential term $(\gamma_{ij}/M_*) \, N_H^2  N^{c}_i N^{c}_j$, leads to a pair of right-handed neutrinos ($N$) or sneutrinos ($\widetilde N$) respectively with equal decay width given by  
\begin{eqnarray}\label{gammaN}
\Gamma_N &=& \Gamma(\theta_H \rightarrow N N) = \Gamma(S \rightarrow \widetilde N \widetilde N)  \\
&=& \frac{M_H }{8 \pi } \left( \frac{M_N}{v_H} \right)^2 \left( 1- \frac{4 M_N^2}{M_H^2} \right)^{1/2}, \qquad
\end{eqnarray}
provided the lightest right-handed neutrino mass $M_N$ satisfies the kinematic condition $M_H \geq 2 M_N$. This channel is crucial for generating the baryon asymmetry via leptogenesis, which requires us to consider $M_H \simeq 2 M_N$. The required kinematic suppression, in turn, makes the decay width of this channel $\Gamma_N$ suppressed compared to $\Gamma_{h,H}$. We further require $M_N \approx 10 \, T_r$
to suppress the washout effects.}}
 {{With $H = 3 \, \Gamma_\text{inf}$, we define the reheat temperature $T_r$ in terms of the inflaton decay width $\Gamma_\text{inf}$,
\begin{equation}\label{tr}
T_r=\left(\dfrac{90}{\pi^{2}g_{*}}\right)^{1/4}\sqrt{\Gamma_\text{inf} \, m_{P}},
\end{equation}
where $g_{*}= 228.75$ for MSSM and $\Gamma_\text{inf}  \simeq \Gamma_h + \Gamma_H$. }}
\begin{table}[t]
\centering
\renewcommand{\arraystretch}{1.0}
\setlength{\tabcolsep}{10pt}
\begin{tabular}{|c|c|c|}
\hline
\textbf{Parameter} & \textbf{Benchmark 1} & \textbf{Benchmark 2}  \\
\hline
$G\mu_\text{cs}$            & $1.3 \times 10^{-7}$ & $1.0 \times 10^{-6}$  \\
$M_\xi$ [GeV]            & $1.8 \times 10^{15}$ & $4.9 \times 10^{15}$ \\
$m_{3/2}$ [TeV]           & $10$              & $10$              \\
$m_{\text{inf}}$  [GeV]        & $5\times10^{9}$             & $5\times10^{9}$              \\
$v_{\phi}$ [GeV]                 & $10^{16}$               & $2 \times 10^{16}$              \\
$M_H$   [GeV]          & $4 \times 10^{10}$ & $4 \times 10^{10}$\\
$T_r$ [GeV]           & $2 \times 10^{9}$ & $2 \times 10^{9}$ \\
$M_N$ [GeV]         & $2 \times 10^{10}$ & $2 \times 10^{10}$  \\
$x_0$       & $1.008 \, (1.007) $ & $1.023 \, (1.02)$  \\
$x_e$      & $1$ & $1$    \\
$n_s$               & $0.965 \, (0.9743)$             & $0.965 (0.9743)$             \\
$r$                 & $9 \times 10^{-14}$ & $5 \times 10^{-12}$  \\
$N_0$          &  $50.6$      &  $51.2$       \\
$\kappa$       &  $10^{-4}$   &  $10^{-4}$    \\
$\kappa_{SS}$  & $1$   & $1$   \\
$\xi$  & $0.2-0.3$   & $0.2-0.3$   \\
$a$  & $0.3 \, (0.2)$   & $9.8 \, (7.0)$   \\
$\gamma_h$  & $2$   & $2$   \\
$\gamma_H$  & $1$   & $1$   \\
$\kappa_S$  & $0.014 \, (0.009)$   & $0.017 \, (0.012)$  \\
\hline
\end{tabular}
\caption{Benchmark points for inflationary and reheating parameters consistent with metastable cosmic strings, CMB observations, and successful leptogenesis ($n_L/s \simeq 2.5 \times 10^{-10}$).}
\label{assign3}
\end{table}

Table~\ref{assign3} summarizes two representative benchmark points that accommodate metastable cosmic strings, are consistent with current CMB data, and yield successful non-thermal leptogenesis. The corresponding predictions for the scalar spectral index are aligned with the central values from Planck 2018 ($n_s = 0.965$) \cite{Planck:2018vyg, Planck:2018jri} and the recent ACT DR6 datasets ($n_s = 0.9743$) \cite{ACT:2025tim}. The baryon asymmetry of the universe is explained through a non-thermal leptogenesis scenario, generating the required lepton asymmetry $n_L/s \simeq 2.5 \times 10^{-10}$ \cite{Planck:2018vyg}.

In supersymmetric hybrid inflation, preheating is typically suppressed unless both the inflaton and the waterfall fields couple to an additional scalar sector. In such cases, efficient preheating may occur if the inflaton $S$ possesses relatively large couplings to those scalars \cite{Garcia-Bellido:1997hex}. In our framework, the electroweak Higgs doublets $h_u,\,h_d$ and the Higgs pair $H,\,\bar{H}$ could in principle play this role. However, efficient preheating would require either $\lambda_h \gg \kappa$ or $\lambda_H \gg \kappa$. Since in our benchmark scenario we consider $\lambda_h \sim \lambda_H \sim \kappa \sim 10^{-4}$, non-perturbative effects from preheating are expected to be strongly suppressed. Moreover, preheating into fermionic channels is generically inefficient due to Pauli blocking, and thus does not alter this conclusion.

\section{Constraints on Current-carrying Metastable Cosmic Strings}
\label{CSs}
 {{The strength of the string's gravitational interaction is expressed in terms of the dimensionless string tension, $G\mu_\text{cs}$, where $G=1/8\pi m_P^2$ and $\mu_\text{cs}$ is mass per unit length of the string. The CMB bound on the standard CS tension is \cite{Planck:2018vyg, Planck:2018jri},
\begin{equation}
G\mu_\text{cs}\lesssim 1.3 \times 10^{-7}.
\end{equation}
The quantity $\mu_\text{cs}$, can be written in terms of the $U(1)_R \times U(1)_{{B-L}}$ gauge symmetry breaking scale $M_{\xi}$ as
$\mu_\text{cs} \simeq 2\pi M_{\xi}^2$. In the standard scenario involving metastable cosmic strings (CSc), one typically obtains $M_{\xi} \lesssim 1.8 \times 10^{15}$~GeV corresponding to $G\mu_\text{cs} \lesssim 1.3 \times 10^{-7}$,  which is compatible with a metastable cosmic string network, as discussed in  \cite{Buchmuller:2019gfy}. This possibility not only circumvents the CMB bound on $G\mu_\text{cs}$, it can also evade other bounds coming from LIGO O3 \cite{KAGRA:2021kbb}. In the case of current-carrying cosmic strings considered in Ref.~\cite{Afzal:2023kqs}, the LIGO O3 bounds can be evaded even with $G\mu_\text{cs} \sim  10^{-6}$ corresponding to $M_{\xi} \sim 4.9 \times 10^{15}$~GeV.}}

 {{The metastable string network decays via the Schwinger production of monopole-antimonopole pairs with a rate per string unit length of \cite{Monin:2009ch,Buchmuller:2019gfy,PhysRevD.78.065048,PhysRevD.79.123519}
\begin{equation}
    \Gamma_d = \frac{\mu_{cs}}{2\pi} \exp(-\pi \kappa_\text{cs}),
\end{equation}
where $\kappa_\text{cs}$ quantifies the metastability of CSs network and is given by
\begin{eqnarray}
\kappa^{1/2}_\text{cs} &\approx & \frac{M_{m}}{\sqrt{\mu_\text{cs}}} \sim \frac{4\pi M_X}{g_m^2}\frac{1}{\sqrt{2\pi}\,v_{\text{str}}}.
\end{eqnarray}
Here $M_{m}$ is the monopole mass and $M_X = \sqrt{2/3}\,(g_m v_m)$ is the gauge boson mass associated with the GUT symmetry breaking responsible for monopole production.  The parameters $v_m$ and $v_{\text{str}}$ represent the scales for the production of monopoles and CSs, respectively. The metastable CSs can explain the NANOGrav observations for $\sqrt{\kappa_\text{cs}} \sim 8$ (see~\Cref{fig:PSmetacs}), which implies,
\begin{eqnarray}
 \frac{ v_m }{ v_{\text{str}}} \sim 2\, g_m, 
\end{eqnarray}
where $g_m = g_4(M_X) = g_3(M_X)$ with $\upsilon_m = 2\,M/(3\,\xi)$ and $v_{\text{str}} = M_{\xi}$. Therefore, we obtain, 
\begin{eqnarray}
\frac{ v_m }{ v_{\text{str}}} \sim \frac{2}{\sqrt{ \frac{4}{3}-9 \,\xi^2}}, 
\end{eqnarray}
\begin{figure}[t]
\centering
\includegraphics[width=0.85\linewidth]{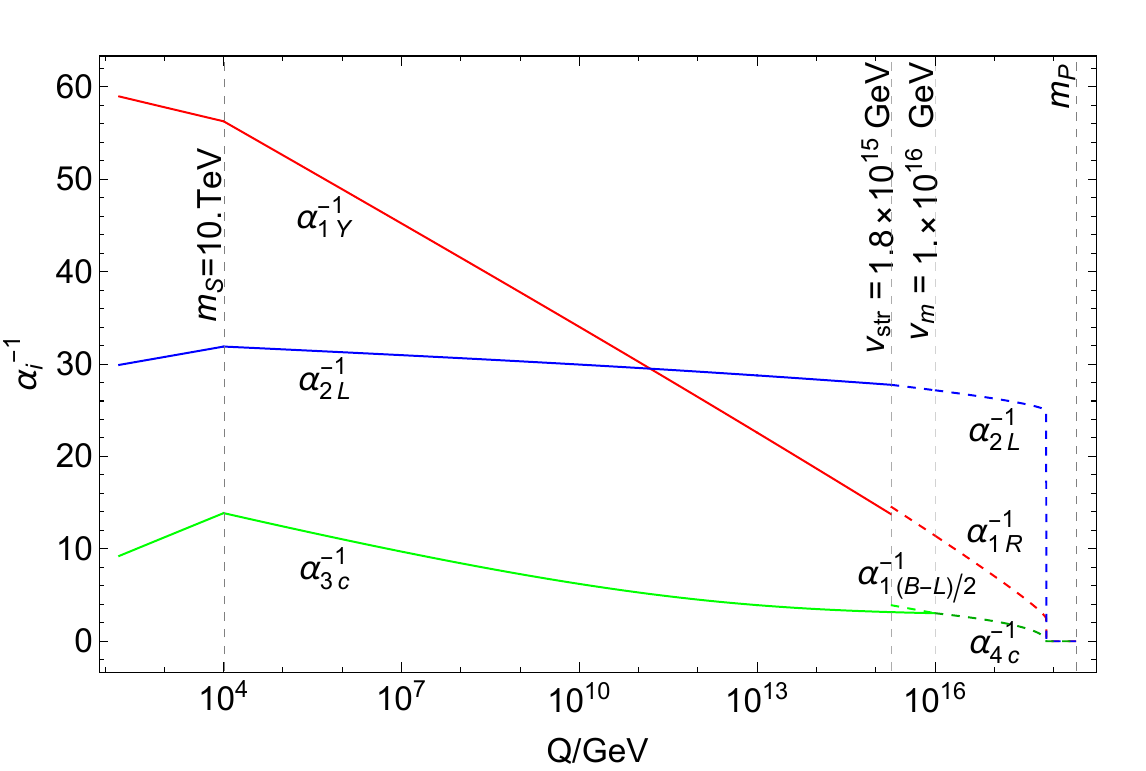}
\caption{\label{gce2}Two loop RG evolution of the gauge couplings for the symmetries $SU(3)_c\times SU(2)_L \times U(1)_Y$, $U(1)_{B-L}$ and $SU(4)_c\times SU(2)_L \times U(1)_R$ with an additional 15-plet and a pair of $H,\bar{H}$. With this additional content, we obtain $g_m \sim 2$.}
\end{figure}
\begin{figure}[t]
\centering
\includegraphics[width=0.85\linewidth]{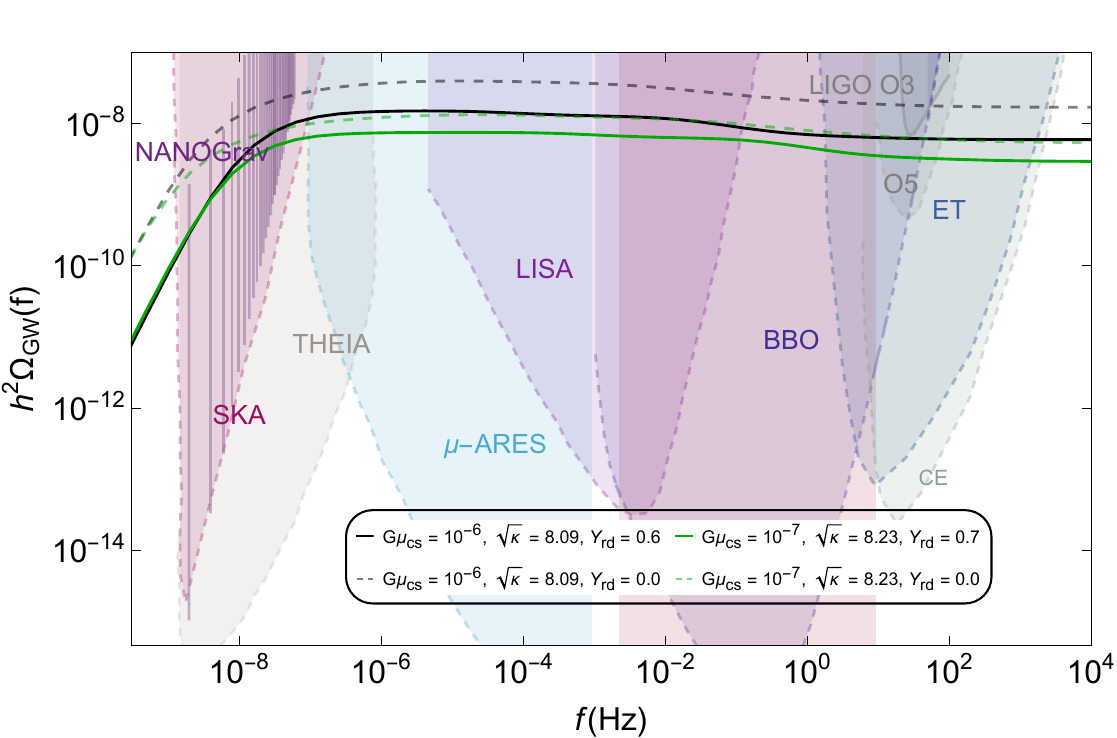}
\caption{The GW spectra of the current carrying CSs for the benchmark points in~\Cref{assign3}. This not only explains the NANOGrav 15-year dataset for $G\mu_\text{cs}\simeq 10^{-6}\,(10^{-7})$ with $\sqrt{\kappa_\text{cs}}\simeq8.09\,(8.23)$ but is also consistent with the LIGO O3 run. The standard scenario has also been presented with the dashed lines. The colored shaded regions indicate the sensitivity curves of present (solid boundaries) LIGO O3 \cite{KAGRA:2021kbb}, NANOGrav \cite{NANOGrav:2023gor}, and future (dashed boundaries) SKA \cite{Smits:2008cf}, THEIA \cite{Garcia-Bellido:2021zgu}, LISA \cite{amaroseoane2017laser}, ET \cite{Punturo:2010zz}, BBO \cite{Corbin:2005ny}, $\mu$-ARES \cite{Sesana:2019vho} and CE \cite{Reitze:2019iox} experiments.}
\label{fig:PSmetacs}
\end{figure} 
implying $g_m \sim 1.4\,(1.0)$ for $\xi \sim 0.3\,(0.2)$. Thus, the existence of metastable CSs compared to account for the NANOGrav results \cite{NANOGrav:2023hvm}. With the minimal matter content described in \Cref{assign1}, we obtain $g_m \sim 1$ as shown in \Cref{gce}. The evolution of gauge couplings under this scenario is illustrated in \Cref{gce2}.
For the benchmark points in \Cref{assign3}, with $G\mu_\text{cs}\simeq 10^{-6}\,(10^{-7})$ and $\sqrt{\kappa_\text{cs}}\simeq 8.09\,(8.23)$, the GW spectrum for the current-carrying CSs is shown in \Cref{fig:PSmetacs}. We follow the dynamics given in \cite{Rybak:2022sbo, Rybak:2024our} for the calculation of current-carrying CSs. The right handed neutrino zero modes residing on the CSs constitute a chiral current, and the major contribution arises from the cusps. In \Cref{fig:PSmetacs}, we have also included the standard case of metastable CSs, and one can see that the GUT scale CSs ($G\mu\simeq 10^{-6}$) are in tension with the LIGO O3 run. But if they are current carrying, as in our model, they not only evade the LIGO O3 bound, but they are consistent with the NANOGrav 15-year dataset as well. Similar results for metastable CSs have previously been presented in \cite{Afzal:2023kqs}.}}
\FloatBarrier

\section{Conclusions }\label{con}
We have shown that superheavy metastable strings can evade inflation in a supersymmetric model based on the gauge symmetry $SU(4)_c \times SU(2)_L \times U(1)_R$.  {{The strings described here are current-carrying due to the presence of right handed neutrino zero modes, and they produce a GW spectrum that appears to be compatible with the recent pulsar timing array experiments and the LIGO O3 run.}} Measurements in other frequency ranges should provide additional tests of the model. Moreover, our considerations here can be extended to the breaking chain
$SO(10) \rightarrow SU(3)_c \times SU(2)_L \times U(1)_Y \times U(1)_{\chi} \rightarrow SU(3)_c \times SU(2)_L \times U(1)_Y$, previously discussed in \cite{Afzal:2022vjx,Maji:2025thf}. The first breaking produces the $SO(10)$ GUT monopole as well as the ‘$\chi$’ monopole, with the latter being the analog of the ‘red’ monopole in our $SU(4)_c \times SU(2)_L \times U(1)_R$ model. Additional care is required to make sure that the symmetry-breaking scales are compatible with the NANOGrav and LIGO-VIRGO measurements.
\section*{Acknowledgments}
Q.S thanks George Lazarides, Amit Tiwari, Rinku Maji, Shaikh Saad, Ahmad Moursy, Steve King, and Stefan Antusch for discussions related to unified theories, monopoles, strings and gravitational waves.

\section*{DATA AVAILABILITY}
The data that support the findings of this article are openly available at Ref.~\cite{datta}.

\FloatBarrier
\bibliographystyle{apsrev4-1}
\bibliography{bibliography}

\begin{thebibliography}{65}%
\makeatletter
\providecommand \@ifxundefined [1]{%
 \@ifx{#1\undefined}
}%
\providecommand \@ifnum [1]{%
 \ifnum #1\expandafter \@firstoftwo
 \else \expandafter \@secondoftwo
 \fi
}%
\providecommand \@ifx [1]{%
 \ifx #1\expandafter \@firstoftwo
 \else \expandafter \@secondoftwo
 \fi
}%
\providecommand \natexlab [1]{#1}%
\providecommand \enquote  [1]{``#1''}%
\providecommand \bibnamefont  [1]{#1}%
\providecommand \bibfnamefont [1]{#1}%
\providecommand \citenamefont [1]{#1}%
\providecommand \href@noop [0]{\@secondoftwo}%
\providecommand \href [0]{\begingroup \@sanitize@url \@href}%
\providecommand \@href[1]{\@@startlink{#1}\@@href}%
\providecommand \@@href[1]{\endgroup#1\@@endlink}%
\providecommand \@sanitize@url [0]{\catcode `\\12\catcode `\$12\catcode
  `\&12\catcode `\#12\catcode `\^12\catcode `\_12\catcode `\%12\relax}%
\providecommand \@@startlink[1]{}%
\providecommand \@@endlink[0]{}%
\providecommand \url  [0]{\begingroup\@sanitize@url \@url }%
\providecommand \@url [1]{\endgroup\@href {#1}{\urlprefix }}%
\providecommand \urlprefix  [0]{URL }%
\providecommand \Eprint [0]{\href }%
\providecommand \doibase [0]{http://dx.doi.org/}%
\providecommand \selectlanguage [0]{\@gobble}%
\providecommand \bibinfo  [0]{\@secondoftwo}%
\providecommand \bibfield  [0]{\@secondoftwo}%
\providecommand \translation [1]{[#1]}%
\providecommand \BibitemOpen [0]{}%
\providecommand \bibitemStop [0]{}%
\providecommand \bibitemNoStop [0]{.\EOS\space}%
\providecommand \EOS [0]{\spacefactor3000\relax}%
\providecommand \BibitemShut  [1]{\csname bibitem#1\endcsname}%
\let\auto@bib@innerbib\@empty
\bibitem [{\citenamefont {Agazie}\ \emph {et~al.}(2023)\citenamefont {Agazie}
  \emph {et~al.}}]{NANOGrav:2023gor}%
  \BibitemOpen
  \bibfield  {author} {\bibinfo {author} {\bibfnamefont {G.}~\bibnamefont
  {Agazie}} \emph {et~al.} (\bibinfo {collaboration} {NANOGrav}),\ }\href
  {\doibase 10.3847/2041-8213/acdac6} {\bibfield  {journal} {\bibinfo
  {journal} {Astrophys. J. Lett.}\ }\textbf {\bibinfo {volume} {951}},\
  \bibinfo {pages} {L8} (\bibinfo {year} {2023})},\ \Eprint
  {http://arxiv.org/abs/2306.16213} {arXiv:2306.16213 [astro-ph.HE]}
  \BibitemShut {NoStop}%
\bibitem [{\citenamefont {Afzal}\ \emph {et~al.}(2023)\citenamefont {Afzal}
  \emph {et~al.}}]{NANOGrav:2023hvm}%
  \BibitemOpen
  \bibfield  {author} {\bibinfo {author} {\bibfnamefont {A.}~\bibnamefont
  {Afzal}} \emph {et~al.} (\bibinfo {collaboration} {NANOGrav}),\ }\href
  {\doibase 10.3847/2041-8213/acdc91} {\bibfield  {journal} {\bibinfo
  {journal} {Astrophys. J. Lett.}\ }\textbf {\bibinfo {volume} {951}},\
  \bibinfo {pages} {L11} (\bibinfo {year} {2023})},\ \Eprint
  {http://arxiv.org/abs/2306.16219} {arXiv:2306.16219 [astro-ph.HE]}
  \BibitemShut {NoStop}%
\bibitem [{\citenamefont {Antoniadis}\ \emph {et~al.}(2023)\citenamefont
  {Antoniadis} \emph {et~al.}}]{EPTA:2023fyk}%
  \BibitemOpen
  \bibfield  {author} {\bibinfo {author} {\bibfnamefont {J.}~\bibnamefont
  {Antoniadis}} \emph {et~al.} (\bibinfo {collaboration} {EPTA}),\ }\href@noop
  {} {\  (\bibinfo {year} {2023})},\ \Eprint {http://arxiv.org/abs/2306.16214}
  {arXiv:2306.16214 [astro-ph.HE]} \BibitemShut {NoStop}%
\bibitem [{\citenamefont {Reardon}\ \emph {et~al.}(2023)\citenamefont {Reardon}
  \emph {et~al.}}]{Reardon:2023gzh}%
  \BibitemOpen
  \bibfield  {author} {\bibinfo {author} {\bibfnamefont {D.~J.}\ \bibnamefont
  {Reardon}} \emph {et~al.},\ }\href {\doibase 10.3847/2041-8213/acdd02}
  {\bibfield  {journal} {\bibinfo  {journal} {Astrophys. J. Lett.}\ }\textbf
  {\bibinfo {volume} {951}},\ \bibinfo {pages} {L6} (\bibinfo {year} {2023})},\
  \Eprint {http://arxiv.org/abs/2306.16215} {arXiv:2306.16215 [astro-ph.HE]}
  \BibitemShut {NoStop}%
\bibitem [{\citenamefont {Xu}\ \emph {et~al.}(2023)\citenamefont {Xu} \emph
  {et~al.}}]{Xu:2023wog}%
  \BibitemOpen
  \bibfield  {author} {\bibinfo {author} {\bibfnamefont {H.}~\bibnamefont {Xu}}
  \emph {et~al.},\ }\href {\doibase 10.1088/1674-4527/acdfa5} {\bibfield
  {journal} {\bibinfo  {journal} {Res. Astron. Astrophys.}\ }\textbf {\bibinfo
  {volume} {23}},\ \bibinfo {pages} {075024} (\bibinfo {year} {2023})},\
  \Eprint {http://arxiv.org/abs/2306.16216} {arXiv:2306.16216 [astro-ph.HE]}
  \BibitemShut {NoStop}%
\bibitem [{\citenamefont {Buchmuller}\ \emph
  {et~al.}(2020{\natexlab{a}})\citenamefont {Buchmuller}, \citenamefont
  {Domcke},\ and\ \citenamefont {Schmitz}}]{Buchmuller:2020lbh}%
  \BibitemOpen
  \bibfield  {author} {\bibinfo {author} {\bibfnamefont {W.}~\bibnamefont
  {Buchmuller}}, \bibinfo {author} {\bibfnamefont {V.}~\bibnamefont {Domcke}},
  \ and\ \bibinfo {author} {\bibfnamefont {K.}~\bibnamefont {Schmitz}},\ }\href
  {\doibase 10.1016/j.physletb.2020.135914} {\bibfield  {journal} {\bibinfo
  {journal} {Phys. Lett. B}\ }\textbf {\bibinfo {volume} {811}},\ \bibinfo
  {pages} {135914} (\bibinfo {year} {2020}{\natexlab{a}})},\ \Eprint
  {http://arxiv.org/abs/2009.10649} {arXiv:2009.10649 [astro-ph.CO]}
  \BibitemShut {NoStop}%
\bibitem [{\citenamefont {Buchmuller}\ \emph {et~al.}(2021)\citenamefont
  {Buchmuller}, \citenamefont {Domcke},\ and\ \citenamefont
  {Schmitz}}]{Buchmuller:2021mbb}%
  \BibitemOpen
  \bibfield  {author} {\bibinfo {author} {\bibfnamefont {W.}~\bibnamefont
  {Buchmuller}}, \bibinfo {author} {\bibfnamefont {V.}~\bibnamefont {Domcke}},
  \ and\ \bibinfo {author} {\bibfnamefont {K.}~\bibnamefont {Schmitz}},\ }\href
  {\doibase 10.1088/1475-7516/2021/12/006} {\bibfield  {journal} {\bibinfo
  {journal} {JCAP}\ }\textbf {\bibinfo {volume} {12}},\ \bibinfo {pages} {006}
  (\bibinfo {year} {2021})},\ \Eprint {http://arxiv.org/abs/2107.04578}
  {arXiv:2107.04578 [hep-ph]} \BibitemShut {NoStop}%
\bibitem [{\citenamefont {Lazarides}\ \emph
  {et~al.}(2023{\natexlab{a}})\citenamefont {Lazarides}, \citenamefont {Maji},
  \citenamefont {Moursy},\ and\ \citenamefont {Shafi}}]{Lazarides:2023rqf}%
  \BibitemOpen
  \bibfield  {author} {\bibinfo {author} {\bibfnamefont {G.}~\bibnamefont
  {Lazarides}}, \bibinfo {author} {\bibfnamefont {R.}~\bibnamefont {Maji}},
  \bibinfo {author} {\bibfnamefont {A.}~\bibnamefont {Moursy}}, \ and\ \bibinfo
  {author} {\bibfnamefont {Q.}~\bibnamefont {Shafi}},\ }\href@noop {} {\
  (\bibinfo {year} {2023}{\natexlab{a}})},\ \Eprint
  {http://arxiv.org/abs/2308.07094} {arXiv:2308.07094 [hep-ph]} \BibitemShut
  {NoStop}%
\bibitem [{\citenamefont {Buchmuller}\ \emph {et~al.}(2023)\citenamefont
  {Buchmuller}, \citenamefont {Domcke},\ and\ \citenamefont
  {Schmitz}}]{Buchmuller:2023aus}%
  \BibitemOpen
  \bibfield  {author} {\bibinfo {author} {\bibfnamefont {W.}~\bibnamefont
  {Buchmuller}}, \bibinfo {author} {\bibfnamefont {V.}~\bibnamefont {Domcke}},
  \ and\ \bibinfo {author} {\bibfnamefont {K.}~\bibnamefont {Schmitz}},\
  }\href@noop {} {\  (\bibinfo {year} {2023})},\ \Eprint
  {http://arxiv.org/abs/2307.04691} {arXiv:2307.04691 [hep-ph]} \BibitemShut
  {NoStop}%
\bibitem [{\citenamefont {Antusch}\ \emph
  {et~al.}(2023{\natexlab{a}})\citenamefont {Antusch}, \citenamefont {Hinze},
  \citenamefont {Saad},\ and\ \citenamefont {Steiner}}]{Antusch:2023zjk}%
  \BibitemOpen
  \bibfield  {author} {\bibinfo {author} {\bibfnamefont {S.}~\bibnamefont
  {Antusch}}, \bibinfo {author} {\bibfnamefont {K.}~\bibnamefont {Hinze}},
  \bibinfo {author} {\bibfnamefont {S.}~\bibnamefont {Saad}}, \ and\ \bibinfo
  {author} {\bibfnamefont {J.}~\bibnamefont {Steiner}},\ }\href@noop {} {\
  (\bibinfo {year} {2023}{\natexlab{a}})},\ \Eprint
  {http://arxiv.org/abs/2307.04595} {arXiv:2307.04595 [hep-ph]} \BibitemShut
  {NoStop}%
\bibitem [{\citenamefont {Ahmed}\ \emph
  {et~al.}(2023{\natexlab{a}})\citenamefont {Ahmed}, \citenamefont {Rehman},\
  and\ \citenamefont {Zubair}}]{Ahmed:2023rky}%
  \BibitemOpen
  \bibfield  {author} {\bibinfo {author} {\bibfnamefont {W.}~\bibnamefont
  {Ahmed}}, \bibinfo {author} {\bibfnamefont {M.~U.}\ \bibnamefont {Rehman}}, \
  and\ \bibinfo {author} {\bibfnamefont {U.}~\bibnamefont {Zubair}},\
  }\href@noop {} {\  (\bibinfo {year} {2023}{\natexlab{a}})},\ \Eprint
  {http://arxiv.org/abs/2308.09125} {arXiv:2308.09125 [hep-ph]} \BibitemShut
  {NoStop}%
\bibitem [{\citenamefont {Lazarides}\ \emph
  {et~al.}(2023{\natexlab{b}})\citenamefont {Lazarides}, \citenamefont {Maji},\
  and\ \citenamefont {Shafi}}]{Lazarides:2023ksx}%
  \BibitemOpen
  \bibfield  {author} {\bibinfo {author} {\bibfnamefont {G.}~\bibnamefont
  {Lazarides}}, \bibinfo {author} {\bibfnamefont {R.}~\bibnamefont {Maji}}, \
  and\ \bibinfo {author} {\bibfnamefont {Q.}~\bibnamefont {Shafi}},\
  }\href@noop {} {\  (\bibinfo {year} {2023}{\natexlab{b}})},\ \Eprint
  {http://arxiv.org/abs/2306.17788} {arXiv:2306.17788 [hep-ph]} \BibitemShut
  {NoStop}%
\bibitem [{\citenamefont {Lazarides}\ \emph
  {et~al.}(2023{\natexlab{c}})\citenamefont {Lazarides}, \citenamefont
  {Shafi},\ and\ \citenamefont {Tiwari}}]{Lazarides:2023iim}%
  \BibitemOpen
  \bibfield  {author} {\bibinfo {author} {\bibfnamefont {G.}~\bibnamefont
  {Lazarides}}, \bibinfo {author} {\bibfnamefont {Q.}~\bibnamefont {Shafi}}, \
  and\ \bibinfo {author} {\bibfnamefont {A.}~\bibnamefont {Tiwari}},\ }\href
  {\doibase 10.1007/JHEP05(2023)119} {\bibfield  {journal} {\bibinfo  {journal}
  {JHEP}\ }\textbf {\bibinfo {volume} {05}},\ \bibinfo {pages} {119} (\bibinfo
  {year} {2023}{\natexlab{c}})},\ \Eprint {http://arxiv.org/abs/2303.15159}
  {arXiv:2303.15159 [hep-ph]} \BibitemShut {NoStop}%
\bibitem [{\citenamefont {Dvali}\ \emph {et~al.}(1994)\citenamefont {Dvali},
  \citenamefont {Shafi},\ and\ \citenamefont {Schaefer}}]{Dvali:1994ms}%
  \BibitemOpen
  \bibfield  {author} {\bibinfo {author} {\bibfnamefont {G.~R.}\ \bibnamefont
  {Dvali}}, \bibinfo {author} {\bibfnamefont {Q.}~\bibnamefont {Shafi}}, \ and\
  \bibinfo {author} {\bibfnamefont {R.~K.}\ \bibnamefont {Schaefer}},\ }\href
  {\doibase 10.1103/PhysRevLett.73.1886} {\bibfield  {journal} {\bibinfo
  {journal} {Phys. Rev. Lett.}\ }\textbf {\bibinfo {volume} {73}},\ \bibinfo
  {pages} {1886} (\bibinfo {year} {1994})},\ \Eprint
  {http://arxiv.org/abs/hep-ph/9406319} {arXiv:hep-ph/9406319} \BibitemShut
  {NoStop}%
\bibitem [{\citenamefont {Copeland}\ \emph {et~al.}(1994)\citenamefont
  {Copeland}, \citenamefont {Liddle}, \citenamefont {Lyth}, \citenamefont
  {Stewart},\ and\ \citenamefont {Wands}}]{Copeland:1994vg}%
  \BibitemOpen
  \bibfield  {author} {\bibinfo {author} {\bibfnamefont {E.~J.}\ \bibnamefont
  {Copeland}}, \bibinfo {author} {\bibfnamefont {A.~R.}\ \bibnamefont
  {Liddle}}, \bibinfo {author} {\bibfnamefont {D.~H.}\ \bibnamefont {Lyth}},
  \bibinfo {author} {\bibfnamefont {E.~D.}\ \bibnamefont {Stewart}}, \ and\
  \bibinfo {author} {\bibfnamefont {D.}~\bibnamefont {Wands}},\ }\href
  {\doibase 10.1103/PhysRevD.49.6410} {\bibfield  {journal} {\bibinfo
  {journal} {Phys. Rev. D}\ }\textbf {\bibinfo {volume} {49}},\ \bibinfo
  {pages} {6410} (\bibinfo {year} {1994})},\ \Eprint
  {http://arxiv.org/abs/astro-ph/9401011} {arXiv:astro-ph/9401011} \BibitemShut
  {NoStop}%
\bibitem [{\citenamefont {Senoguz}\ and\ \citenamefont
  {Shafi}(2005)}]{Senoguz:2004vu}%
  \BibitemOpen
  \bibfield  {author} {\bibinfo {author} {\bibfnamefont {V.~N.}\ \bibnamefont
  {Senoguz}}\ and\ \bibinfo {author} {\bibfnamefont {Q.}~\bibnamefont
  {Shafi}},\ }\href {\doibase 10.1103/PhysRevD.71.043514} {\bibfield  {journal}
  {\bibinfo  {journal} {Phys. Rev. D}\ }\textbf {\bibinfo {volume} {71}},\
  \bibinfo {pages} {043514} (\bibinfo {year} {2005})},\ \Eprint
  {http://arxiv.org/abs/hep-ph/0412102} {arXiv:hep-ph/0412102} \BibitemShut
  {NoStop}%
\bibitem [{\citenamefont {Rehman}\ \emph
  {et~al.}(2010{\natexlab{a}})\citenamefont {Rehman}, \citenamefont {Shafi},\
  and\ \citenamefont {Wickman}}]{Rehman:2009nq}%
  \BibitemOpen
  \bibfield  {author} {\bibinfo {author} {\bibfnamefont {M.~U.}\ \bibnamefont
  {Rehman}}, \bibinfo {author} {\bibfnamefont {Q.}~\bibnamefont {Shafi}}, \
  and\ \bibinfo {author} {\bibfnamefont {J.~R.}\ \bibnamefont {Wickman}},\
  }\href {\doibase 10.1016/j.physletb.2009.12.010} {\bibfield  {journal}
  {\bibinfo  {journal} {Phys. Lett. B}\ }\textbf {\bibinfo {volume} {683}},\
  \bibinfo {pages} {191} (\bibinfo {year} {2010}{\natexlab{a}})},\ \Eprint
  {http://arxiv.org/abs/0908.3896} {arXiv:0908.3896 [hep-ph]} \BibitemShut
  {NoStop}%
\bibitem [{\citenamefont {Buchmuller}\ \emph {et~al.}(2000)\citenamefont
  {Buchmuller}, \citenamefont {Covi},\ and\ \citenamefont
  {Delepine}}]{Buchmuller:2000zm}%
  \BibitemOpen
  \bibfield  {author} {\bibinfo {author} {\bibfnamefont {W.}~\bibnamefont
  {Buchmuller}}, \bibinfo {author} {\bibfnamefont {L.}~\bibnamefont {Covi}}, \
  and\ \bibinfo {author} {\bibfnamefont {D.}~\bibnamefont {Delepine}},\ }\href
  {\doibase 10.1016/S0370-2693(00)01005-4} {\bibfield  {journal} {\bibinfo
  {journal} {Phys. Lett. B}\ }\textbf {\bibinfo {volume} {491}},\ \bibinfo
  {pages} {183} (\bibinfo {year} {2000})},\ \Eprint
  {http://arxiv.org/abs/hep-ph/0006168} {arXiv:hep-ph/0006168} \BibitemShut
  {NoStop}%
\bibitem [{\citenamefont {Bastero-Gil}\ \emph {et~al.}(2007)\citenamefont
  {Bastero-Gil}, \citenamefont {King},\ and\ \citenamefont
  {Shafi}}]{Bastero-Gil:2006zpr}%
  \BibitemOpen
  \bibfield  {author} {\bibinfo {author} {\bibfnamefont {M.}~\bibnamefont
  {Bastero-Gil}}, \bibinfo {author} {\bibfnamefont {S.~F.}\ \bibnamefont
  {King}}, \ and\ \bibinfo {author} {\bibfnamefont {Q.}~\bibnamefont {Shafi}},\
  }\href {\doibase 10.1016/j.physletb.2006.06.085} {\bibfield  {journal}
  {\bibinfo  {journal} {Phys. Lett. B}\ }\textbf {\bibinfo {volume} {651}},\
  \bibinfo {pages} {345} (\bibinfo {year} {2007})},\ \Eprint
  {http://arxiv.org/abs/hep-ph/0604198} {arXiv:hep-ph/0604198} \BibitemShut
  {NoStop}%
\bibitem [{\citenamefont {ur~Rehman}\ \emph {et~al.}(2007)\citenamefont
  {ur~Rehman}, \citenamefont {Senoguz},\ and\ \citenamefont
  {Shafi}}]{urRehman:2006hu}%
  \BibitemOpen
  \bibfield  {author} {\bibinfo {author} {\bibfnamefont {M.}~\bibnamefont
  {ur~Rehman}}, \bibinfo {author} {\bibfnamefont {V.~N.}\ \bibnamefont
  {Senoguz}}, \ and\ \bibinfo {author} {\bibfnamefont {Q.}~\bibnamefont
  {Shafi}},\ }\href {\doibase 10.1103/PhysRevD.75.043522} {\bibfield  {journal}
  {\bibinfo  {journal} {Phys. Rev. D}\ }\textbf {\bibinfo {volume} {75}},\
  \bibinfo {pages} {043522} (\bibinfo {year} {2007})},\ \Eprint
  {http://arxiv.org/abs/hep-ph/0612023} {arXiv:hep-ph/0612023} \BibitemShut
  {NoStop}%
\bibitem [{\citenamefont {Rehman}\ \emph
  {et~al.}(2010{\natexlab{b}})\citenamefont {Rehman}, \citenamefont {Shafi},\
  and\ \citenamefont {Wickman}}]{Rehman:2009yj}%
  \BibitemOpen
  \bibfield  {author} {\bibinfo {author} {\bibfnamefont {M.~U.}\ \bibnamefont
  {Rehman}}, \bibinfo {author} {\bibfnamefont {Q.}~\bibnamefont {Shafi}}, \
  and\ \bibinfo {author} {\bibfnamefont {J.~R.}\ \bibnamefont {Wickman}},\
  }\href {\doibase 10.1016/j.physletb.2010.03.072} {\bibfield  {journal}
  {\bibinfo  {journal} {Phys. Lett. B}\ }\textbf {\bibinfo {volume} {688}},\
  \bibinfo {pages} {75} (\bibinfo {year} {2010}{\natexlab{b}})},\ \Eprint
  {http://arxiv.org/abs/0912.4737} {arXiv:0912.4737 [hep-ph]} \BibitemShut
  {NoStop}%
\bibitem [{\citenamefont {Shafi}\ and\ \citenamefont
  {Wickman}(2011)}]{Shafi:2010jr}%
  \BibitemOpen
  \bibfield  {author} {\bibinfo {author} {\bibfnamefont {Q.}~\bibnamefont
  {Shafi}}\ and\ \bibinfo {author} {\bibfnamefont {J.~R.}\ \bibnamefont
  {Wickman}},\ }\href {\doibase 10.1016/j.physletb.2011.01.002} {\bibfield
  {journal} {\bibinfo  {journal} {Phys. Lett. B}\ }\textbf {\bibinfo {volume}
  {696}},\ \bibinfo {pages} {438} (\bibinfo {year} {2011})},\ \Eprint
  {http://arxiv.org/abs/1009.5340} {arXiv:1009.5340 [hep-ph]} \BibitemShut
  {NoStop}%
\bibitem [{\citenamefont {Rehman}\ \emph {et~al.}(2011)\citenamefont {Rehman},
  \citenamefont {Shafi},\ and\ \citenamefont {Wickman}}]{Rehman:2010wm}%
  \BibitemOpen
  \bibfield  {author} {\bibinfo {author} {\bibfnamefont {M.~U.}\ \bibnamefont
  {Rehman}}, \bibinfo {author} {\bibfnamefont {Q.}~\bibnamefont {Shafi}}, \
  and\ \bibinfo {author} {\bibfnamefont {J.~R.}\ \bibnamefont {Wickman}},\
  }\href {\doibase 10.1103/PhysRevD.83.067304} {\bibfield  {journal} {\bibinfo
  {journal} {Phys. Rev. D}\ }\textbf {\bibinfo {volume} {83}},\ \bibinfo
  {pages} {067304} (\bibinfo {year} {2011})},\ \Eprint
  {http://arxiv.org/abs/1012.0309} {arXiv:1012.0309 [astro-ph.CO]} \BibitemShut
  {NoStop}%
\bibitem [{\citenamefont {Civiletti}\ \emph {et~al.}(2011)\citenamefont
  {Civiletti}, \citenamefont {Rehman}, \citenamefont {Shafi},\ and\
  \citenamefont {Wickman}}]{Civiletti:2011qg}%
  \BibitemOpen
  \bibfield  {author} {\bibinfo {author} {\bibfnamefont {M.}~\bibnamefont
  {Civiletti}}, \bibinfo {author} {\bibfnamefont {M.~U.}\ \bibnamefont
  {Rehman}}, \bibinfo {author} {\bibfnamefont {Q.}~\bibnamefont {Shafi}}, \
  and\ \bibinfo {author} {\bibfnamefont {J.~R.}\ \bibnamefont {Wickman}},\
  }\href {\doibase 10.1103/PhysRevD.84.103505} {\bibfield  {journal} {\bibinfo
  {journal} {Phys. Rev. D}\ }\textbf {\bibinfo {volume} {84}},\ \bibinfo
  {pages} {103505} (\bibinfo {year} {2011})},\ \Eprint
  {http://arxiv.org/abs/1104.4143} {arXiv:1104.4143 [astro-ph.CO]} \BibitemShut
  {NoStop}%
\bibitem [{\citenamefont {Buchm\"uller}\ \emph {et~al.}(2014)\citenamefont
  {Buchm\"uller}, \citenamefont {Domcke}, \citenamefont {Kamada},\ and\
  \citenamefont {Schmitz}}]{Buchmuller:2014epa}%
  \BibitemOpen
  \bibfield  {author} {\bibinfo {author} {\bibfnamefont {W.}~\bibnamefont
  {Buchm\"uller}}, \bibinfo {author} {\bibfnamefont {V.}~\bibnamefont
  {Domcke}}, \bibinfo {author} {\bibfnamefont {K.}~\bibnamefont {Kamada}}, \
  and\ \bibinfo {author} {\bibfnamefont {K.}~\bibnamefont {Schmitz}},\ }\href
  {\doibase 10.1088/1475-7516/2014/07/054} {\bibfield  {journal} {\bibinfo
  {journal} {JCAP}\ }\textbf {\bibinfo {volume} {07}},\ \bibinfo {pages} {054}
  (\bibinfo {year} {2014})},\ \Eprint {http://arxiv.org/abs/1404.1832}
  {arXiv:1404.1832 [hep-ph]} \BibitemShut {NoStop}%
\bibitem [{\citenamefont {Buchmuller}(2021)}]{Buchmuller:2021dtt}%
  \BibitemOpen
  \bibfield  {author} {\bibinfo {author} {\bibfnamefont {W.}~\bibnamefont
  {Buchmuller}},\ }\href {\doibase 10.1007/JHEP04(2021)168} {\bibfield
  {journal} {\bibinfo  {journal} {JHEP}\ }\textbf {\bibinfo {volume} {04}},\
  \bibinfo {pages} {168} (\bibinfo {year} {2021})},\ \Eprint
  {http://arxiv.org/abs/2102.08923} {arXiv:2102.08923 [hep-ph]} \BibitemShut
  {NoStop}%
\bibitem [{\citenamefont {Ahmed}\ \emph {et~al.}(2022)\citenamefont {Ahmed},
  \citenamefont {Junaid}, \citenamefont {Nasri},\ and\ \citenamefont
  {Zubair}}]{Ahmed:2022rwy}%
  \BibitemOpen
  \bibfield  {author} {\bibinfo {author} {\bibfnamefont {W.}~\bibnamefont
  {Ahmed}}, \bibinfo {author} {\bibfnamefont {M.}~\bibnamefont {Junaid}},
  \bibinfo {author} {\bibfnamefont {S.}~\bibnamefont {Nasri}}, \ and\ \bibinfo
  {author} {\bibfnamefont {U.}~\bibnamefont {Zubair}},\ }\href {\doibase
  10.1103/PhysRevD.105.115008} {\bibfield  {journal} {\bibinfo  {journal}
  {Phys. Rev. D}\ }\textbf {\bibinfo {volume} {105}},\ \bibinfo {pages}
  {115008} (\bibinfo {year} {2022})},\ \Eprint
  {http://arxiv.org/abs/2202.06216} {arXiv:2202.06216 [hep-ph]} \BibitemShut
  {NoStop}%
\bibitem [{\citenamefont {Ahmed}\ \emph
  {et~al.}(2023{\natexlab{b}})\citenamefont {Ahmed}, \citenamefont {Moosa},
  \citenamefont {Munir},\ and\ \citenamefont {Zubair}}]{Ahmed:2022wed}%
  \BibitemOpen
  \bibfield  {author} {\bibinfo {author} {\bibfnamefont {W.}~\bibnamefont
  {Ahmed}}, \bibinfo {author} {\bibfnamefont {M.}~\bibnamefont {Moosa}},
  \bibinfo {author} {\bibfnamefont {S.}~\bibnamefont {Munir}}, \ and\ \bibinfo
  {author} {\bibfnamefont {U.}~\bibnamefont {Zubair}},\ }\href {\doibase
  10.1007/JHEP05(2023)011} {\bibfield  {journal} {\bibinfo  {journal} {JHEP}\
  }\textbf {\bibinfo {volume} {05}},\ \bibinfo {pages} {011} (\bibinfo {year}
  {2023}{\natexlab{b}})},\ \Eprint {http://arxiv.org/abs/2208.11888}
  {arXiv:2208.11888 [hep-ph]} \BibitemShut {NoStop}%
\bibitem [{\citenamefont {Masoud}\ \emph {et~al.}(2021)\citenamefont {Masoud},
  \citenamefont {Rehman},\ and\ \citenamefont {Shafi}}]{Masoud:2021prr}%
  \BibitemOpen
  \bibfield  {author} {\bibinfo {author} {\bibfnamefont {M.~A.}\ \bibnamefont
  {Masoud}}, \bibinfo {author} {\bibfnamefont {M.~U.}\ \bibnamefont {Rehman}},
  \ and\ \bibinfo {author} {\bibfnamefont {Q.}~\bibnamefont {Shafi}},\ }\href
  {\doibase 10.1088/1475-7516/2021/11/022} {\bibfield  {journal} {\bibinfo
  {journal} {JCAP}\ }\textbf {\bibinfo {volume} {11}},\ \bibinfo {pages} {022}
  (\bibinfo {year} {2021})},\ \Eprint {http://arxiv.org/abs/2107.09689}
  {arXiv:2107.09689 [hep-ph]} \BibitemShut {NoStop}%
\bibitem [{\citenamefont {Okada}\ and\ \citenamefont
  {Shafi}(2017)}]{Okada:2015vka}%
  \BibitemOpen
  \bibfield  {author} {\bibinfo {author} {\bibfnamefont {N.}~\bibnamefont
  {Okada}}\ and\ \bibinfo {author} {\bibfnamefont {Q.}~\bibnamefont {Shafi}},\
  }\href {\doibase 10.1016/j.physletb.2017.11.015} {\bibfield  {journal}
  {\bibinfo  {journal} {Phys. Lett. B}\ }\textbf {\bibinfo {volume} {775}},\
  \bibinfo {pages} {348} (\bibinfo {year} {2017})},\ \Eprint
  {http://arxiv.org/abs/1506.01410} {arXiv:1506.01410 [hep-ph]} \BibitemShut
  {NoStop}%
\bibitem [{\citenamefont {Rehman}\ \emph {et~al.}(2017)\citenamefont {Rehman},
  \citenamefont {Shafi},\ and\ \citenamefont {Vardag}}]{Rehman:2017gkm}%
  \BibitemOpen
  \bibfield  {author} {\bibinfo {author} {\bibfnamefont {M.~U.}\ \bibnamefont
  {Rehman}}, \bibinfo {author} {\bibfnamefont {Q.}~\bibnamefont {Shafi}}, \
  and\ \bibinfo {author} {\bibfnamefont {F.~K.}\ \bibnamefont {Vardag}},\
  }\href {\doibase 10.1103/PhysRevD.96.063527} {\bibfield  {journal} {\bibinfo
  {journal} {Phys. Rev. D}\ }\textbf {\bibinfo {volume} {96}},\ \bibinfo
  {pages} {063527} (\bibinfo {year} {2017})},\ \Eprint
  {http://arxiv.org/abs/1705.03693} {arXiv:1705.03693 [hep-ph]} \BibitemShut
  {NoStop}%
\bibitem [{\citenamefont {Okada}\ and\ \citenamefont
  {Shafi}(2018)}]{Okada:2017rbf}%
  \BibitemOpen
  \bibfield  {author} {\bibinfo {author} {\bibfnamefont {N.}~\bibnamefont
  {Okada}}\ and\ \bibinfo {author} {\bibfnamefont {Q.}~\bibnamefont {Shafi}},\
  }\href {\doibase 10.1016/j.physletb.2018.10.057} {\bibfield  {journal}
  {\bibinfo  {journal} {Phys. Lett. B}\ }\textbf {\bibinfo {volume} {787}},\
  \bibinfo {pages} {141} (\bibinfo {year} {2018})},\ \Eprint
  {http://arxiv.org/abs/1709.04610} {arXiv:1709.04610 [hep-ph]} \BibitemShut
  {NoStop}%
\bibitem [{\citenamefont {Ahmed}\ \emph {et~al.}(2021)\citenamefont {Ahmed},
  \citenamefont {Karozas},\ and\ \citenamefont {Leontaris}}]{Ahmed:2021dvo}%
  \BibitemOpen
  \bibfield  {author} {\bibinfo {author} {\bibfnamefont {W.}~\bibnamefont
  {Ahmed}}, \bibinfo {author} {\bibfnamefont {A.}~\bibnamefont {Karozas}}, \
  and\ \bibinfo {author} {\bibfnamefont {G.~K.}\ \bibnamefont {Leontaris}},\
  }\href@noop {} {\bibfield  {journal} {\bibinfo  {journal} {Phys.Rev.D}\
  }\textbf {\bibinfo {volume} {104}},\ \bibinfo {pages} {055025} (\bibinfo
  {year} {2021})},\ \Eprint {http://arxiv.org/abs/2104.04328} {arXiv:2104.04328
  [hep-ph]} \BibitemShut {NoStop}%
\bibitem [{\citenamefont {Afzal}\ \emph {et~al.}(2022)\citenamefont {Afzal},
  \citenamefont {Ahmed}, \citenamefont {Rehman},\ and\ \citenamefont
  {Shafi}}]{Afzal:2022vjx}%
  \BibitemOpen
  \bibfield  {author} {\bibinfo {author} {\bibfnamefont {A.}~\bibnamefont
  {Afzal}}, \bibinfo {author} {\bibfnamefont {W.}~\bibnamefont {Ahmed}},
  \bibinfo {author} {\bibfnamefont {M.~U.}\ \bibnamefont {Rehman}}, \ and\
  \bibinfo {author} {\bibfnamefont {Q.}~\bibnamefont {Shafi}},\ }\href
  {\doibase 10.1103/PhysRevD.105.103539} {\bibfield  {journal} {\bibinfo
  {journal} {Phys. Rev. D}\ }\textbf {\bibinfo {volume} {105}},\ \bibinfo
  {pages} {103539} (\bibinfo {year} {2022})},\ \Eprint
  {http://arxiv.org/abs/2202.07386} {arXiv:2202.07386 [hep-ph]} \BibitemShut
  {NoStop}%
\bibitem [{\citenamefont {Lazarides}\ and\ \citenamefont
  {Shafi}(2019)}]{Lazarides:2019xai}%
  \BibitemOpen
  \bibfield  {author} {\bibinfo {author} {\bibfnamefont {G.}~\bibnamefont
  {Lazarides}}\ and\ \bibinfo {author} {\bibfnamefont {Q.}~\bibnamefont
  {Shafi}},\ }\href {\doibase 10.1007/JHEP10(2019)193} {\bibfield  {journal}
  {\bibinfo  {journal} {JHEP}\ }\textbf {\bibinfo {volume} {10}},\ \bibinfo
  {pages} {193} (\bibinfo {year} {2019})},\ \Eprint
  {http://arxiv.org/abs/1904.06880} {arXiv:1904.06880 [hep-ph]} \BibitemShut
  {NoStop}%
\bibitem [{\citenamefont {Lazarides}\ \emph {et~al.}(2021)\citenamefont
  {Lazarides}, \citenamefont {Rehman}, \citenamefont {Shafi},\ and\
  \citenamefont {Vardag}}]{Lazarides:2020zof}%
  \BibitemOpen
  \bibfield  {author} {\bibinfo {author} {\bibfnamefont {G.}~\bibnamefont
  {Lazarides}}, \bibinfo {author} {\bibfnamefont {M.~U.}\ \bibnamefont
  {Rehman}}, \bibinfo {author} {\bibfnamefont {Q.}~\bibnamefont {Shafi}}, \
  and\ \bibinfo {author} {\bibfnamefont {F.~K.}\ \bibnamefont {Vardag}},\
  }\href {\doibase 10.1103/PhysRevD.103.035033} {\bibfield  {journal} {\bibinfo
   {journal} {Phys. Rev. D}\ }\textbf {\bibinfo {volume} {103}},\ \bibinfo
  {pages} {035033} (\bibinfo {year} {2021})},\ \Eprint
  {http://arxiv.org/abs/2007.01474} {arXiv:2007.01474 [hep-ph]} \BibitemShut
  {NoStop}%
\bibitem [{\citenamefont {Ahmed}\ \emph {et~al.}(2024)\citenamefont {Ahmed},
  \citenamefont {Mehmood}, \citenamefont {Rehman},\ and\ \citenamefont
  {Zubair}}]{Ahmed:2024iyd}%
  \BibitemOpen
  \bibfield  {author} {\bibinfo {author} {\bibfnamefont {W.}~\bibnamefont
  {Ahmed}}, \bibinfo {author} {\bibfnamefont {M.}~\bibnamefont {Mehmood}},
  \bibinfo {author} {\bibfnamefont {M.~U.}\ \bibnamefont {Rehman}}, \ and\
  \bibinfo {author} {\bibfnamefont {U.}~\bibnamefont {Zubair}},\ }\href@noop {}
  {\  (\bibinfo {year} {2024})},\ \Eprint {http://arxiv.org/abs/2404.06008}
  {arXiv:2404.06008 [hep-ph]} \BibitemShut {NoStop}%
\bibitem [{\citenamefont {Aad}\ \emph {et~al.}(2023)\citenamefont {Aad} \emph
  {et~al.}}]{ATLAS:2022pib}%
  \BibitemOpen
  \bibfield  {author} {\bibinfo {author} {\bibfnamefont {G.}~\bibnamefont
  {Aad}} \emph {et~al.} (\bibinfo {collaboration} {ATLAS}),\ }\href {\doibase
  10.1007/JHEP06(2023)158} {\bibfield  {journal} {\bibinfo  {journal} {JHEP}\
  }\textbf {\bibinfo {volume} {2306}},\ \bibinfo {pages} {158} (\bibinfo {year}
  {2023})},\ \Eprint {http://arxiv.org/abs/2205.06013} {arXiv:2205.06013
  [hep-ex]} \BibitemShut {NoStop}%
\bibitem [{\citenamefont {Antusch}\ \emph
  {et~al.}(2023{\natexlab{b}})\citenamefont {Antusch}, \citenamefont {Hinze},
  \citenamefont {Saad},\ and\ \citenamefont {Steiner}}]{Antusch:2023mxx}%
  \BibitemOpen
  \bibfield  {author} {\bibinfo {author} {\bibfnamefont {S.}~\bibnamefont
  {Antusch}}, \bibinfo {author} {\bibfnamefont {K.}~\bibnamefont {Hinze}},
  \bibinfo {author} {\bibfnamefont {S.}~\bibnamefont {Saad}}, \ and\ \bibinfo
  {author} {\bibfnamefont {J.}~\bibnamefont {Steiner}},\ }\href@noop {} {\
  (\bibinfo {year} {2023}{\natexlab{b}})},\ \Eprint
  {http://arxiv.org/abs/2308.11705} {arXiv:2308.11705 [hep-ph]} \BibitemShut
  {NoStop}%
\bibitem [{\citenamefont {Khalil}\ \emph {et~al.}(2011)\citenamefont {Khalil},
  \citenamefont {Rehman}, \citenamefont {Shafi},\ and\ \citenamefont
  {Zaakouk}}]{Khalil:2010cp}%
  \BibitemOpen
  \bibfield  {author} {\bibinfo {author} {\bibfnamefont {S.}~\bibnamefont
  {Khalil}}, \bibinfo {author} {\bibfnamefont {M.~U.}\ \bibnamefont {Rehman}},
  \bibinfo {author} {\bibfnamefont {Q.}~\bibnamefont {Shafi}}, \ and\ \bibinfo
  {author} {\bibfnamefont {E.~A.}\ \bibnamefont {Zaakouk}},\ }\href {\doibase
  10.1103/PhysRevD.83.063522} {\bibfield  {journal} {\bibinfo  {journal} {Phys.
  Rev. D}\ }\textbf {\bibinfo {volume} {83}},\ \bibinfo {pages} {063522}
  (\bibinfo {year} {2011})},\ \Eprint {http://arxiv.org/abs/1010.3657}
  {arXiv:1010.3657 [hep-ph]} \BibitemShut {NoStop}%
\bibitem [{\citenamefont {Senoguz}\ and\ \citenamefont
  {Shafi}(2008)}]{Senoguz:2008nok}%
  \BibitemOpen
  \bibfield  {author} {\bibinfo {author} {\bibfnamefont {V.~N.}\ \bibnamefont
  {Senoguz}}\ and\ \bibinfo {author} {\bibfnamefont {Q.}~\bibnamefont
  {Shafi}},\ }\href {\doibase 10.1016/j.physletb.2008.08.017} {\bibfield
  {journal} {\bibinfo  {journal} {Phys. Lett. B}\ }\textbf {\bibinfo {volume}
  {668}},\ \bibinfo {pages} {6} (\bibinfo {year} {2008})},\ \Eprint
  {http://arxiv.org/abs/0806.2798} {arXiv:0806.2798 [hep-ph]} \BibitemShut
  {NoStop}%
\bibitem [{\citenamefont {Senoguz}\ and\ \citenamefont
  {Shafi}(2004)}]{Senoguz:2004ky}%
  \BibitemOpen
  \bibfield  {author} {\bibinfo {author} {\bibfnamefont {V.~N.}\ \bibnamefont
  {Senoguz}}\ and\ \bibinfo {author} {\bibfnamefont {Q.}~\bibnamefont
  {Shafi}},\ }\href {\doibase 10.1016/j.physletb.2004.05.077} {\bibfield
  {journal} {\bibinfo  {journal} {Phys. Lett. B}\ }\textbf {\bibinfo {volume}
  {596}},\ \bibinfo {pages} {8} (\bibinfo {year} {2004})},\ \Eprint
  {http://arxiv.org/abs/hep-ph/0403294} {arXiv:hep-ph/0403294} \BibitemShut
  {NoStop}%
\bibitem [{\citenamefont {Dvali}\ \emph {et~al.}(1998)\citenamefont {Dvali},
  \citenamefont {Lazarides},\ and\ \citenamefont {Shafi}}]{Dvali:1997uq}%
  \BibitemOpen
  \bibfield  {author} {\bibinfo {author} {\bibfnamefont {G.~R.}\ \bibnamefont
  {Dvali}}, \bibinfo {author} {\bibfnamefont {G.}~\bibnamefont {Lazarides}}, \
  and\ \bibinfo {author} {\bibfnamefont {Q.}~\bibnamefont {Shafi}},\ }\href
  {\doibase 10.1016/S0370-2693(98)00145-2} {\bibfield  {journal} {\bibinfo
  {journal} {Phys. Lett. B}\ }\textbf {\bibinfo {volume} {424}},\ \bibinfo
  {pages} {259} (\bibinfo {year} {1998})},\ \Eprint
  {http://arxiv.org/abs/hep-ph/9710314} {arXiv:hep-ph/9710314} \BibitemShut
  {NoStop}%
\bibitem [{\citenamefont {Lazarides}\ and\ \citenamefont
  {Vlachos}(1998)}]{Lazarides:1998qx}%
  \BibitemOpen
  \bibfield  {author} {\bibinfo {author} {\bibfnamefont {G.}~\bibnamefont
  {Lazarides}}\ and\ \bibinfo {author} {\bibfnamefont {N.~D.}\ \bibnamefont
  {Vlachos}},\ }\href {\doibase 10.1016/S0370-2693(98)01126-5} {\bibfield
  {journal} {\bibinfo  {journal} {Phys. Lett. B}\ }\textbf {\bibinfo {volume}
  {441}},\ \bibinfo {pages} {46} (\bibinfo {year} {1998})},\ \Eprint
  {http://arxiv.org/abs/hep-ph/9807253} {arXiv:hep-ph/9807253} \BibitemShut
  {NoStop}%
\bibitem [{\citenamefont {Aghanim}\ \emph {et~al.}(2020)\citenamefont {Aghanim}
  \emph {et~al.}}]{Planck:2018vyg}%
  \BibitemOpen
  \bibfield  {author} {\bibinfo {author} {\bibfnamefont {N.}~\bibnamefont
  {Aghanim}} \emph {et~al.} (\bibinfo {collaboration} {Planck}),\ }\href
  {\doibase 10.1051/0004-6361/201833910} {\bibfield  {journal} {\bibinfo
  {journal} {Astron. Astrophys.}\ }\textbf {\bibinfo {volume} {641}},\ \bibinfo
  {pages} {A6} (\bibinfo {year} {2020})},\ \bibinfo {note} {[Erratum:
  Astron.Astrophys. 652, C4 (2021)]},\ \Eprint
  {http://arxiv.org/abs/1807.06209} {arXiv:1807.06209 [astro-ph.CO]}
  \BibitemShut {NoStop}%
\bibitem [{\citenamefont {Akrami}\ \emph {et~al.}(2020)\citenamefont {Akrami}
  \emph {et~al.}}]{Planck:2018jri}%
  \BibitemOpen
  \bibfield  {author} {\bibinfo {author} {\bibfnamefont {Y.}~\bibnamefont
  {Akrami}} \emph {et~al.} (\bibinfo {collaboration} {Planck}),\ }\href
  {\doibase 10.1051/0004-6361/201833887} {\bibfield  {journal} {\bibinfo
  {journal} {Astron. Astrophys.}\ }\textbf {\bibinfo {volume} {641}},\ \bibinfo
  {pages} {A10} (\bibinfo {year} {2020})},\ \Eprint
  {http://arxiv.org/abs/1807.06211} {arXiv:1807.06211 [astro-ph.CO]}
  \BibitemShut {NoStop}%
\bibitem [{\citenamefont {Calabrese}\ \emph {et~al.}(2025)\citenamefont
  {Calabrese} \emph {et~al.}}]{ACT:2025tim}%
  \BibitemOpen
  \bibfield  {author} {\bibinfo {author} {\bibfnamefont {E.}~\bibnamefont
  {Calabrese}} \emph {et~al.} (\bibinfo {collaboration} {ACT}),\ }\href@noop {}
  {\  (\bibinfo {year} {2025})},\ \Eprint {http://arxiv.org/abs/2503.14454}
  {arXiv:2503.14454 [astro-ph.CO]} \BibitemShut {NoStop}%
\bibitem [{\citenamefont {Garcia-Bellido}\ and\ \citenamefont
  {Linde}(1998)}]{Garcia-Bellido:1997hex}%
  \BibitemOpen
  \bibfield  {author} {\bibinfo {author} {\bibfnamefont {J.}~\bibnamefont
  {Garcia-Bellido}}\ and\ \bibinfo {author} {\bibfnamefont {A.~D.}\
  \bibnamefont {Linde}},\ }\href {\doibase 10.1103/PhysRevD.57.6075} {\bibfield
   {journal} {\bibinfo  {journal} {Phys. Rev. D}\ }\textbf {\bibinfo {volume}
  {57}},\ \bibinfo {pages} {6075} (\bibinfo {year} {1998})},\ \Eprint
  {http://arxiv.org/abs/hep-ph/9711360} {arXiv:hep-ph/9711360} \BibitemShut
  {NoStop}%
\bibitem [{\citenamefont {Buchmuller}\ \emph
  {et~al.}(2020{\natexlab{b}})\citenamefont {Buchmuller}, \citenamefont
  {Domcke}, \citenamefont {Murayama},\ and\ \citenamefont
  {Schmitz}}]{Buchmuller:2019gfy}%
  \BibitemOpen
  \bibfield  {author} {\bibinfo {author} {\bibfnamefont {W.}~\bibnamefont
  {Buchmuller}}, \bibinfo {author} {\bibfnamefont {V.}~\bibnamefont {Domcke}},
  \bibinfo {author} {\bibfnamefont {H.}~\bibnamefont {Murayama}}, \ and\
  \bibinfo {author} {\bibfnamefont {K.}~\bibnamefont {Schmitz}},\ }\href
  {\doibase 10.1016/j.physletb.2020.135764} {\bibfield  {journal} {\bibinfo
  {journal} {Phys. Lett. B}\ }\textbf {\bibinfo {volume} {809}},\ \bibinfo
  {pages} {135764} (\bibinfo {year} {2020}{\natexlab{b}})},\ \Eprint
  {http://arxiv.org/abs/1912.03695} {arXiv:1912.03695 [hep-ph]} \BibitemShut
  {NoStop}%
\bibitem [{\citenamefont {Abbott}\ \emph {et~al.}(2021)\citenamefont {Abbott}
  \emph {et~al.}}]{KAGRA:2021kbb}%
  \BibitemOpen
  \bibfield  {author} {\bibinfo {author} {\bibfnamefont {R.}~\bibnamefont
  {Abbott}} \emph {et~al.} (\bibinfo {collaboration} {KAGRA, Virgo, LIGO
  Scientific}),\ }\href {\doibase 10.1103/PhysRevD.104.022004} {\bibfield
  {journal} {\bibinfo  {journal} {Phys. Rev. D}\ }\textbf {\bibinfo {volume}
  {104}},\ \bibinfo {pages} {022004} (\bibinfo {year} {2021})},\ \Eprint
  {http://arxiv.org/abs/2101.12130} {arXiv:2101.12130 [gr-qc]} \BibitemShut
  {NoStop}%
\bibitem [{\citenamefont {Afzal}\ \emph {et~al.}(2024)\citenamefont {Afzal},
  \citenamefont {Shafi},\ and\ \citenamefont {Tiwari}}]{Afzal:2023kqs}%
  \BibitemOpen
  \bibfield  {author} {\bibinfo {author} {\bibfnamefont {A.}~\bibnamefont
  {Afzal}}, \bibinfo {author} {\bibfnamefont {Q.}~\bibnamefont {Shafi}}, \ and\
  \bibinfo {author} {\bibfnamefont {A.}~\bibnamefont {Tiwari}},\ }\href
  {\doibase 10.1016/j.physletb.2024.138516} {\bibfield  {journal} {\bibinfo
  {journal} {Phys. Lett. B}\ }\textbf {\bibinfo {volume} {850}},\ \bibinfo
  {pages} {138516} (\bibinfo {year} {2024})},\ \Eprint
  {http://arxiv.org/abs/2311.05564} {arXiv:2311.05564 [hep-ph]} \BibitemShut
  {NoStop}%
\bibitem [{\citenamefont {Monin}\ and\ \citenamefont
  {Voloshin}(2010)}]{Monin:2009ch}%
  \BibitemOpen
  \bibfield  {author} {\bibinfo {author} {\bibfnamefont {A.}~\bibnamefont
  {Monin}}\ and\ \bibinfo {author} {\bibfnamefont {M.~B.}\ \bibnamefont
  {Voloshin}},\ }\href {\doibase 10.1134/S1063778810040162} {\bibfield
  {journal} {\bibinfo  {journal} {Phys. Atom. Nucl.}\ }\textbf {\bibinfo
  {volume} {73}},\ \bibinfo {pages} {703} (\bibinfo {year} {2010})},\ \Eprint
  {http://arxiv.org/abs/0902.0407} {arXiv:0902.0407 [hep-th]} \BibitemShut
  {NoStop}%
\bibitem [{\citenamefont {Monin}\ and\ \citenamefont
  {Voloshin}(2008)}]{PhysRevD.78.065048}%
  \BibitemOpen
  \bibfield  {author} {\bibinfo {author} {\bibfnamefont {A.}~\bibnamefont
  {Monin}}\ and\ \bibinfo {author} {\bibfnamefont {M.~B.}\ \bibnamefont
  {Voloshin}},\ }\href {\doibase 10.1103/PhysRevD.78.065048} {\bibfield
  {journal} {\bibinfo  {journal} {Phys. Rev. D}\ }\textbf {\bibinfo {volume}
  {78}},\ \bibinfo {pages} {065048} (\bibinfo {year} {2008})}\BibitemShut
  {NoStop}%
\bibitem [{\citenamefont {Leblond}\ \emph {et~al.}(2009)\citenamefont
  {Leblond}, \citenamefont {Shlaer},\ and\ \citenamefont
  {Siemens}}]{PhysRevD.79.123519}%
  \BibitemOpen
  \bibfield  {author} {\bibinfo {author} {\bibfnamefont {L.}~\bibnamefont
  {Leblond}}, \bibinfo {author} {\bibfnamefont {B.}~\bibnamefont {Shlaer}}, \
  and\ \bibinfo {author} {\bibfnamefont {X.}~\bibnamefont {Siemens}},\ }\href
  {\doibase 10.1103/PhysRevD.79.123519} {\bibfield  {journal} {\bibinfo
  {journal} {Phys. Rev. D}\ }\textbf {\bibinfo {volume} {79}},\ \bibinfo
  {pages} {123519} (\bibinfo {year} {2009})}\BibitemShut {NoStop}%
\bibitem [{\citenamefont {Smits}\ \emph {et~al.}(2009)\citenamefont {Smits},
  \citenamefont {Kramer}, \citenamefont {Stappers}, \citenamefont {Lorimer},
  \citenamefont {Cordes},\ and\ \citenamefont {Faulkner}}]{Smits:2008cf}%
  \BibitemOpen
  \bibfield  {author} {\bibinfo {author} {\bibfnamefont {R.}~\bibnamefont
  {Smits}}, \bibinfo {author} {\bibfnamefont {M.}~\bibnamefont {Kramer}},
  \bibinfo {author} {\bibfnamefont {B.}~\bibnamefont {Stappers}}, \bibinfo
  {author} {\bibfnamefont {D.~R.}\ \bibnamefont {Lorimer}}, \bibinfo {author}
  {\bibfnamefont {J.}~\bibnamefont {Cordes}}, \ and\ \bibinfo {author}
  {\bibfnamefont {A.}~\bibnamefont {Faulkner}},\ }\href {\doibase
  10.1051/0004-6361:200810383} {\bibfield  {journal} {\bibinfo  {journal}
  {Astron. Astrophys.}\ }\textbf {\bibinfo {volume} {493}},\ \bibinfo {pages}
  {1161} (\bibinfo {year} {2009})},\ \Eprint {http://arxiv.org/abs/0811.0211}
  {arXiv:0811.0211 [astro-ph]} \BibitemShut {NoStop}%
\bibitem [{\citenamefont {Garcia-Bellido}\ \emph {et~al.}(2021)\citenamefont
  {Garcia-Bellido}, \citenamefont {Murayama},\ and\ \citenamefont
  {White}}]{Garcia-Bellido:2021zgu}%
  \BibitemOpen
  \bibfield  {author} {\bibinfo {author} {\bibfnamefont {J.}~\bibnamefont
  {Garcia-Bellido}}, \bibinfo {author} {\bibfnamefont {H.}~\bibnamefont
  {Murayama}}, \ and\ \bibinfo {author} {\bibfnamefont {G.}~\bibnamefont
  {White}},\ }\href {\doibase 10.1088/1475-7516/2021/12/023} {\bibfield
  {journal} {\bibinfo  {journal} {JCAP}\ }\textbf {\bibinfo {volume} {12}},\
  \bibinfo {pages} {023} (\bibinfo {year} {2021})},\ \Eprint
  {http://arxiv.org/abs/2104.04778} {arXiv:2104.04778 [hep-ph]} \BibitemShut
  {NoStop}%
\bibitem [{\citenamefont {{Pau Amaro-Seoane, et
  al.}}(2017)}]{amaroseoane2017laser}%
  \BibitemOpen
  \bibfield  {author} {\bibinfo {author} {\bibnamefont {{Pau Amaro-Seoane, et
  al.}}},\ }\href@noop {} {\  (\bibinfo {year} {2017})},\ \Eprint
  {http://arxiv.org/abs/1702.00786} {arXiv:1702.00786 [astro-ph.IM]}
  \BibitemShut {NoStop}%
\bibitem [{\citenamefont {Punturo}\ \emph {et~al.}(2010)\citenamefont {Punturo}
  \emph {et~al.}}]{Punturo:2010zz}%
  \BibitemOpen
  \bibfield  {author} {\bibinfo {author} {\bibfnamefont {M.}~\bibnamefont
  {Punturo}} \emph {et~al.},\ }\href {\doibase 10.1088/0264-9381/27/19/194002}
  {\bibfield  {journal} {\bibinfo  {journal} {Class. Quant. Grav.}\ }\textbf
  {\bibinfo {volume} {27}},\ \bibinfo {pages} {194002} (\bibinfo {year}
  {2010})}\BibitemShut {NoStop}%
\bibitem [{\citenamefont {Corbin}\ and\ \citenamefont
  {Cornish}(2006)}]{Corbin:2005ny}%
  \BibitemOpen
  \bibfield  {author} {\bibinfo {author} {\bibfnamefont {V.}~\bibnamefont
  {Corbin}}\ and\ \bibinfo {author} {\bibfnamefont {N.~J.}\ \bibnamefont
  {Cornish}},\ }\href {\doibase 10.1088/0264-9381/23/7/014} {\bibfield
  {journal} {\bibinfo  {journal} {Class. Quant. Grav.}\ }\textbf {\bibinfo
  {volume} {23}},\ \bibinfo {pages} {2435} (\bibinfo {year} {2006})},\ \Eprint
  {http://arxiv.org/abs/gr-qc/0512039} {arXiv:gr-qc/0512039} \BibitemShut
  {NoStop}%
\bibitem [{\citenamefont {Sesana}\ \emph {et~al.}(2021)\citenamefont {Sesana}
  \emph {et~al.}}]{Sesana:2019vho}%
  \BibitemOpen
  \bibfield  {author} {\bibinfo {author} {\bibfnamefont {A.}~\bibnamefont
  {Sesana}} \emph {et~al.},\ }\href {\doibase 10.1007/s10686-021-09709-9}
  {\bibfield  {journal} {\bibinfo  {journal} {Exper. Astron.}\ }\textbf
  {\bibinfo {volume} {51}},\ \bibinfo {pages} {1333} (\bibinfo {year}
  {2021})},\ \Eprint {http://arxiv.org/abs/1908.11391} {arXiv:1908.11391
  [astro-ph.IM]} \BibitemShut {NoStop}%
\bibitem [{\citenamefont {Reitze}\ \emph {et~al.}(2019)\citenamefont {Reitze}
  \emph {et~al.}}]{Reitze:2019iox}%
  \BibitemOpen
  \bibfield  {author} {\bibinfo {author} {\bibfnamefont {D.}~\bibnamefont
  {Reitze}} \emph {et~al.},\ }\href@noop {} {\bibfield  {journal} {\bibinfo
  {journal} {Bull. Am. Astron. Soc.}\ }\textbf {\bibinfo {volume} {51}},\
  \bibinfo {pages} {035} (\bibinfo {year} {2019})},\ \Eprint
  {http://arxiv.org/abs/1907.04833} {arXiv:1907.04833 [astro-ph.IM]}
  \BibitemShut {NoStop}%
\bibitem [{\citenamefont {Rybak}\ and\ \citenamefont
  {Sousa}(2022)}]{Rybak:2022sbo}%
  \BibitemOpen
  \bibfield  {author} {\bibinfo {author} {\bibfnamefont {I.~Y.}\ \bibnamefont
  {Rybak}}\ and\ \bibinfo {author} {\bibfnamefont {L.}~\bibnamefont {Sousa}},\
  }\href {\doibase 10.1088/1475-7516/2022/11/024} {\bibfield  {journal}
  {\bibinfo  {journal} {JCAP}\ }\textbf {\bibinfo {volume} {11}},\ \bibinfo
  {pages} {024} (\bibinfo {year} {2022})},\ \Eprint
  {http://arxiv.org/abs/2209.01068} {arXiv:2209.01068 [gr-qc]} \BibitemShut
  {NoStop}%
\bibitem [{\citenamefont {Rybak}\ and\ \citenamefont
  {Sousa}(2025)}]{Rybak:2024our}%
  \BibitemOpen
  \bibfield  {author} {\bibinfo {author} {\bibfnamefont {I.~Y.}\ \bibnamefont
  {Rybak}}\ and\ \bibinfo {author} {\bibfnamefont {L.}~\bibnamefont {Sousa}},\
  }\href {\doibase 10.1103/PhysRevD.111.083502} {\bibfield  {journal} {\bibinfo
   {journal} {Phys. Rev. D}\ }\textbf {\bibinfo {volume} {111}},\ \bibinfo
  {pages} {083502} (\bibinfo {year} {2025})},\ \Eprint
  {http://arxiv.org/abs/2412.17154} {arXiv:2412.17154 [astro-ph.CO]}
  \BibitemShut {NoStop}%
\bibitem [{\citenamefont {Maji}\ and\ \citenamefont
  {Shafi}(2025)}]{Maji:2025thf}%
  \BibitemOpen
  \bibfield  {author} {\bibinfo {author} {\bibfnamefont {R.}~\bibnamefont
  {Maji}}\ and\ \bibinfo {author} {\bibfnamefont {Q.}~\bibnamefont {Shafi}},\
  }\href {\doibase 10.1007/JHEP06(2025)217} {\bibfield  {journal} {\bibinfo
  {journal} {JHEP}\ }\textbf {\bibinfo {volume} {06}},\ \bibinfo {pages} {217}
  (\bibinfo {year} {2025})},\ \Eprint {http://arxiv.org/abs/2504.09055}
  {arXiv:2504.09055 [hep-ph]} \BibitemShut {NoStop}%
\bibitem [{dat()}]{datta}%
  \BibitemOpen
  \href@noop {} {}\bibinfo {howpublished}
  {\url{https://github.com/mansoorhep/2308.11410}}\BibitemShut {NoStop}%
\end{thebibliography}%
\end{document}